# Atomic-scale tunable phonon transport at tailored grain boundaries and Their Impact on Thermal Conductivity

Xiaowang Wang[1,7], Chaitanya A. Gadre[1,7], Wanjuan Zou[2], Runqing Yang[3], Bin Xing[2], Christopher Addiego[1], Fenghui Gong[2], Toshihiro Aoki[4], Yujie Quan[3], Wei-Tao Peng[5], Yifeng Huang[1], Chaojie Du[2], Mingjie Xu[4], Xingxu Yan[2], Ruqian Wu[1], Shyue Ping Ong[5], Bolin Liao[3], Penghui Cao[2,6], and Xiaoqing Pan [1,2,4*]

[1] Department of Physics and Astronomy, University of California, Irvine, CA 92697, USA

[2] Department of Materials Science and Engineering, University of California, Irvine, CA 92697, USA

[3] Department of Mechanical Engineering, University of California, Santa Barbara, CA 93106, USA

[4] Irvine Materials Research Institute, University of California-Irvine, Irvine, CA 92697, USA

[5] Department of Nanoengineering Engineering, University of California, San Diego, CA 92093. USA

[6] Department of Mechanical Engineering, University of California, Irvine, CA 92697, USA

[7] These authors contributed equally: Xiaowang Wang and Chaitanya A. Gadre

*Email: xiaoqinp@uci.edu

**Grain boundaries has been of fundamental importance in impacting thermal properties in materials for advancing innovative technologies, as heat carriers such as phonons are impeded by breaking crystal symmetry or periodicity[1–3]. Grain boundaries impede phonon propagation include varying the density of defects, interfaces, and nanostructures[1,3–5] as well as changing composition. Although GBs are diverse, a robust systematic study that links GB structure, nanoscale phonons, and thermal conductivities is missing. Here we reveal from nanoscale structure-phonon mechanisms on how the grain boundary (GB) tilt and twist angles fundamentally drive the changes in atom rearrangements, defect specific vibrational states, and finally macroscopic heat transport at different bicrystal strontium titanate GBs using emerging atomic resolution vibrational spectroscopy[6–8]. The low angle tilt GBs (2°, 6°, 10°) primarily modulate phonon population and mode shift via inter-defect distances, while high angle GBs (22°, 36°) thermal conductivity show limited variation from the extent of GB disorder. In addition, GB twist angles introduce coexisting periodic defects that can modulate thermal characteristics locally around the defect. Our results demonstrate that varying the tilt angle coarsely modifies the phonon population along entire GB while varying the twist angle incurs a finer adjustment at periodic locations on the GB. Our study offers a systematic approach to understanding and manipulating cross-GB thermal transport of arbitrary GBs predictably and precisely[4,9–12].**

**Another version of abstract, feel free to take or leave it**

**Grain boundaries (GBs) strongly influence thermal transport in crystalline solids by disrupting lattice periodicity and scattering phonons[1–3]. Due to the atomic-level disorder and structural complexity, a fundamental understanding of how specific GB geometries regulate nanoscale phonon behavior and macroscopic thermal conductivity has remained elusive [1,3–5]. Here, using emerging atomic-resolution vibrational spectroscopy[6–8], we directly correlate GB structure, defect-specific vibrational states, and thermal transport in bicrystal strontium titanate with controlled tilt and twist angles. The phonon characterizations and thermal conductivity data reveal two distinct regimes, where low-angle tilt GBs (2°, 6°, 10°) substantially modulate phonon populations and mode frequencies, resulting in pronounced changes in thermal conductivity, whereas high-angle tilt GBs (22°, 36°) exhibit weak conductivity variation due to saturated structural disorder and scattering. In contrast, twist GBs introduce periodic defect motifs that only locally tailor phonon transport. Our results suggest tilt and twist angles as complementary knobs for coarse and fine control of phonon propagation and cross-GB thermal transport, providing a predictive framework for thermal engineering in materials[4,9–12].**

Nanoscale structural engineering plays a crucial role in managing phonon transport and the thermal conductivity of materials and has been of great interest in material science research due to its broad applications. Many devices with novel functionalities, such as thermal rectification, thermal switching, thermal cloaking, thermoelectrics, and energy storage require desired thermal conductivity by implementing diverse nanoscale engineering strategies. The vast majority of nansocale devices uses crystaline materials as the fundamental building block, making grain boundaries ubiquitous within those devices.

Thermal conductivity tuning can be achieved by infiltrating, doping, and oxidizing the many material[13,14].In two dimensional materials, thermal tuning is accomplished by strain and isotope engineering. Semiconductors often make use of interlayering different materials and interfaces to scatter thermal carriers to modulate thermal conductivity. Domain walls in ferroelectrics have also been used as efficient phonon scatterers by applying electrical biasing to vary domain wall density. Although low-dimensional materials exhibit high flexibility and are easily tailored[15–18], they are often too delicate and lack the robustness of bulk crystals to be used in many technologically critical devices. Crystaline materials, on the other hand, are often not sensitive to the aformentioned thermal conductivity tuning methods. In order to devise an elegant and methodological way to tune the thermal conductivity of bulk crystals, a robust understanding of GB, a natural feature inherent to crystaline materials that drastically altersits thermal transport properties, is necessary .

Conventionally, altering phonon behavior in crystalline materials involves varying defect density, alloying, or varying grain density[2,5,19]which is a somewhat crude way to tune thermal conductivity and may unintentionally affect desirable, intrinsic properties[4,9,20]. The pattern of how a single GB affects thermal conductivity, however, has not been systematically observed by correlating nanoscale phonons and macroscopic thermal conductivity. Grain boundaries (GBs), formed between two single crystal grains by tilt and twist misalignments[21,22], are atomically well-defined defects that are known to impede cross-GB thermal transport[23–28]. It is well known that GBs can drastically alter a material's thermal conductivity and that different GB show differences in thermal conductivity by up to two orders of magnitude[4,19,23,29,30]. Local GB vibrational studies have been achieved with the advancement of vibrational electron energy loss spectroscopy (EELS)[4,7,28,31–34]. Unveiling phonon transport mechanisms at the atomic scale, however, has also been challenging until recently due to the lack of simultaneous spatial and energy resolutions for probing phonon characteristics within a few angstroms near crystalline imperfections. Previous studies considering detailed atomic structures have mainly been theoretical, while experimental studies have resorted to macroscopic investigations using ensembles of GBs to determine their effect on thermal transport[5,35]. With the advent of vibrational EELS, Hoglund et all explored the local phonon density of states of 10 degree strontium titanate GB[36]. Haas et all explored the local vibrational DOS of silicon GBs[33]. These studies show that vibrational electron energy-loss spectroscopy (EELS) in a scanning transmission electron microscope (STEM) is a suitable method for this study, with atomic spatial resolution and few-meV energy resolution capable of capturing phonon states at an unprecedented scale[2,6,8,37–40]. Even though nanoscale GB phonons and macroscopic GB thermal conductivity have been separately studied, a systematic observation of the link between the two, demonstrating the pattern of GB thermal behavior with misorientation angles, is missing. The Goal of this study is to systematically investigate the nanoscale phonons of a series of tilt GBs and combine the findings with macroscopic and simulated thermal conductivities.

Here, using atomic resolution dark-field vibrational EELS, we map phonon populations and mode mismatch of five tilt and twist configuration GBs. We unveil that increasing the tilt angle between grains drastically lowers the phonon population by lowering the inter-defect distance for low angle tilt GBs. This trend continues until the GB tilt angle increases to 11° , where the effective inter-detect distance cannot be further increased as defects fill up the entire GB. With tilt angle greater than 11° the thermal conductivity continues to decrease with increasing angle, but with much less sensitivity. At high tilt angles, the lack of bulk-like regions and increased structural disorder at the GB, drives a more drastic departure of the high-tilt GB phonons from bulk than that of low-tilt ones. We categorize this as a transition of thermal behavior from small-angle dislocation mediated to large angle disorder mediated regime. At the high-angle regime, we found that density of coincidence lattice sites at the GB play a very negligible role in enhancing thermal conductivity. The introduction of a very small (less than 1°) twist angle on the other hand, produces a long periodic structure that coexists with the tilt-induced GB structure and varying this angle changes the periodicity and length. This continuous, disordered twist GB structure more effectively impedes phonons across its length than a pure tilt structure. As a result, varying the twist angle can fine-tune thermal transport. This suggests that modulating grain tilt and twist angles provides a robust mechanism for the coarse and fine tuning of thermal conductivity. Our results also provide an in-depth understanding of GB phonon transport mechanisms by observing patterns at atomic scale that underpin macroscopic thermal conductivity for crystalline materials with limited defect concentrations.

**Commented [XW1]:** 1. add page number
2. add indentation (4 spaces)
3. 3rd paragraph, first sentence: 通常，通过改变晶体材料中的声子行为涉及改变缺陷密度、合金化或改变晶粒密度 2,5,19 来调节热导，很少有人关注晶界特性对热导的巨大影响.

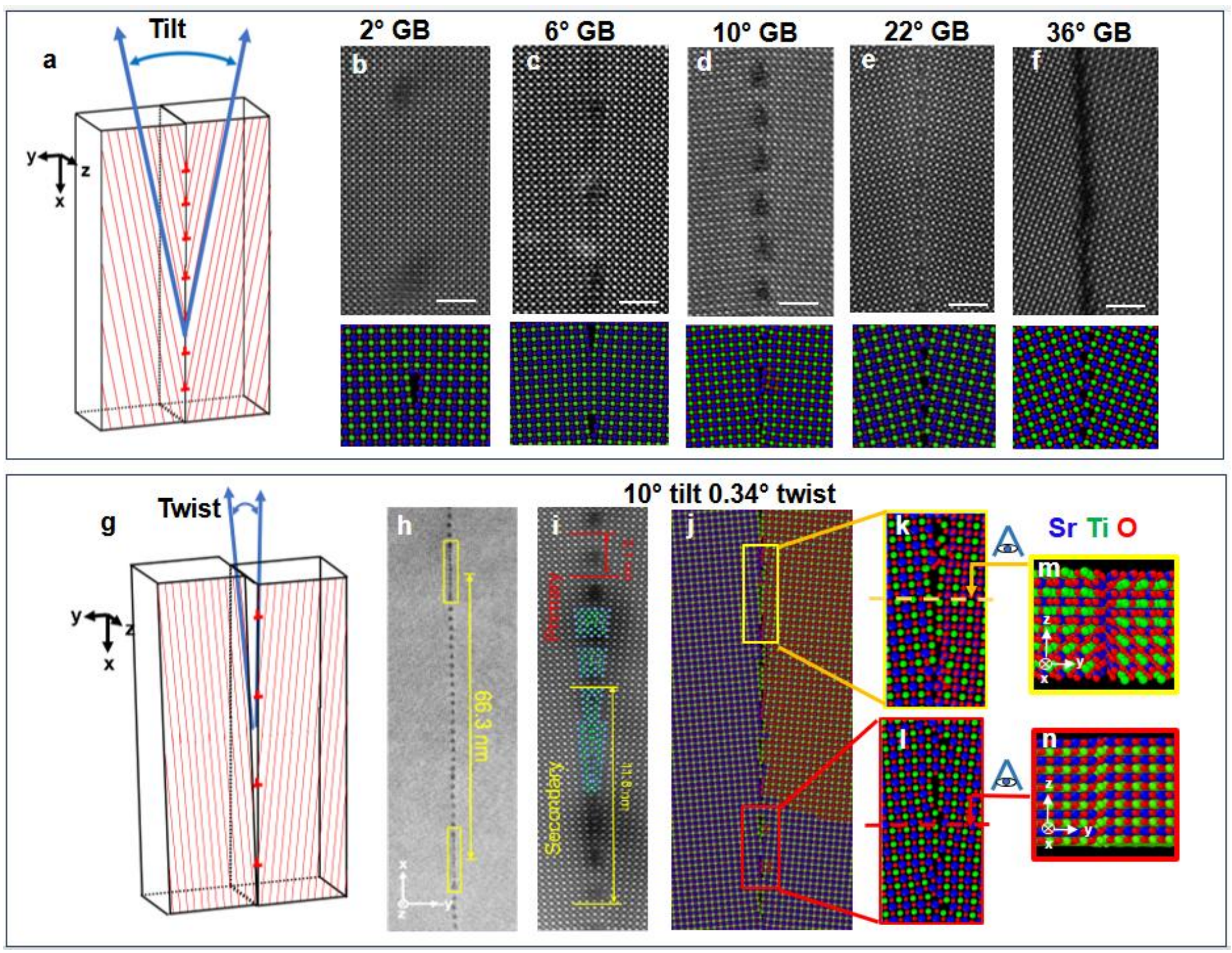

**Figure 1. Atomic structural characterizations of the grain boundaries**. (a) Schematic illustrating symmetric GB tilt rotation. (b-f) HAADF and molecular dynamics (MD) simulation model of 2°, 6°, 10°, 22°, and 36° GB respectively. (g) Schematic illustrating twist rotation. (h) Atomic HAADF image of 10° GB with 0.34° twist, showing twist induced cores with large periodicity. (i) Enlarged atomic image of twist induced cores at 10° GB. (k) MD simulation reproducing coexistence of twist and tilt induced cores. (k) Enlarged view of twist induced structure. (l) Enlarged view of tilt-induced structure. (m) Cross section of twist induced structure with cross sectional plane and viewing direction indicated in (k). (n) Cross section of tilt induced structure with cross sectional plane and viewing direction indicated in (l).

The tilt and twist angles between adjacent grains produce two types of coexisting structures along the GB which we define as tilt-induced and twist-induced defect cores (Fig. 1h). A tilt misalignment is formed by a rotation about the z-axis and produces tilt induced edge dislocation defect cores while a twist misalignment is formed by a rotation about the y-axis producing long-period twist induced screw dislocation defect cores along the GB. Low tilt angles correspond to larger, well-separated cores (Fig. 1b-d) while high tilt angles produce structures that are more contiguous in nature (Fig. 1e-f). Functionally, the twist angle controls the length and periodicity of the twist-induced defect core, which is determined by the periodic misalignment of the atomic planes in the z-direction (Fig. 1 m-n, Extended Fig. 1). Molecular dynamics (MD) using large simulation cells are performed to reproduce the tilt GB periodicities. The results show that both observed and simulated tilt GB periodicity agree with Frank's equation. For twist GBs MD is also useful for reproducing periodicity and providing a 3-dimensional view of the twist inducced defect. The cross-sectional views obtained from MD simulations (Fig. 1m-n) reveal that in the twist-induced core, the SrO plane of one grain is aligned with the $TiO_2$ plane of the other grain, while in the tilt induced core, the respective SrO and $TiO_2$ planes are matched across the grains. The major distinction between the tilt- and twist-induced structures is the presence of bridges (bulk-like structures) between the tilt-induced cores (Fig. 1b-d). The absence of bulk-like bridge structures in the twist-induced core is due to the twist-induced misalignment of the SrO and $TiO_2$ planes (Fig. 1m, n). MD simulations reveal that the length and periodicity of the twist-induced cores are very sensitive to small changes in the twist angle. The proportion of the GB consisting of twist and tilt induced cores monotonically increases with the twist angle. Additionally, the atomic arrangement within each core remains largely unchanged with variations in the twist angle. The way the atomic-scale structure depends on the macroscopically controlled twist and tilt angles forms a robust basis for highly tunable properties. Specifically, variations of the tilt angle correspond to more drastic changes in the GB atomic structure, while twist angle modulations introduce twist-induced defects whose length and periodicity can be adjusted leading to finer adjustments to the GB structure.

Low magnification high-angle annular dark-field (HAADF) image illustrates the twist-induced core in STO, which is created by a minute twist about the y- axis in addition to the tilt-induced dislocation cores owing to a 10° tilt. (Fig. 1h). Tilt-induced cores are defined by local circular regions of lattice disorder with a diameter of 1.2 nm and a periodicity of 2.1 nm. The twist-induced cores include alternating regions of high and low lattice disorder with a period of 66.3 nm and length of 11.8 nm (Fig. 1h,i) owing to a twist angle of 0.34°. A detailed analysis of the tilt and twist induced defect cores is presented in Extended Data Fig. 2a-b and Extended Data Fig. 3, where the unusually long length of the twist-induced core becomes more obvious, and the atomic disorder is more evident. Regions of low disorder such as the bulk-like bridge structures are expected to have a higher number of phonons states that would allow for higher phonon and therefore, thermal transport. Conversely, twist-induced defects with their regions of higher and lower lattice disorder offer no bridge structures. The defect cores of the 22° and 36° GB (Fig. 1e-f) are placed much closer to each other than in the case of the 10° GB, where periodic bulk-like bridges can be seen. Essentially, increasing the tilt angle reduces the periodicity and size of the cores which should greatly affect cross-GB thermal transport properties.

The periodicity of the twist-induced structure must be the periodicity of the planes of the two crystals that intersect due to the twist angle. The twist angle β can be calculated from the period using basic trigonometry: $\beta = \arctan(d/L)$, to be 0.34°, where $d$ is the distance between adjacent (001) planes of STO (3.94 Å), and $L$ is the observed period (see Fig S1. (g). Therefore, a small change in twist angle can induce a remarkable change in the period and thus the concentration of twist-induced cores at the GB. This change in ratio may drastically alter thermal properties of the GBs. This quantitative tunability is important for the ability to accurately adjust nanoscale phonon transport properties.

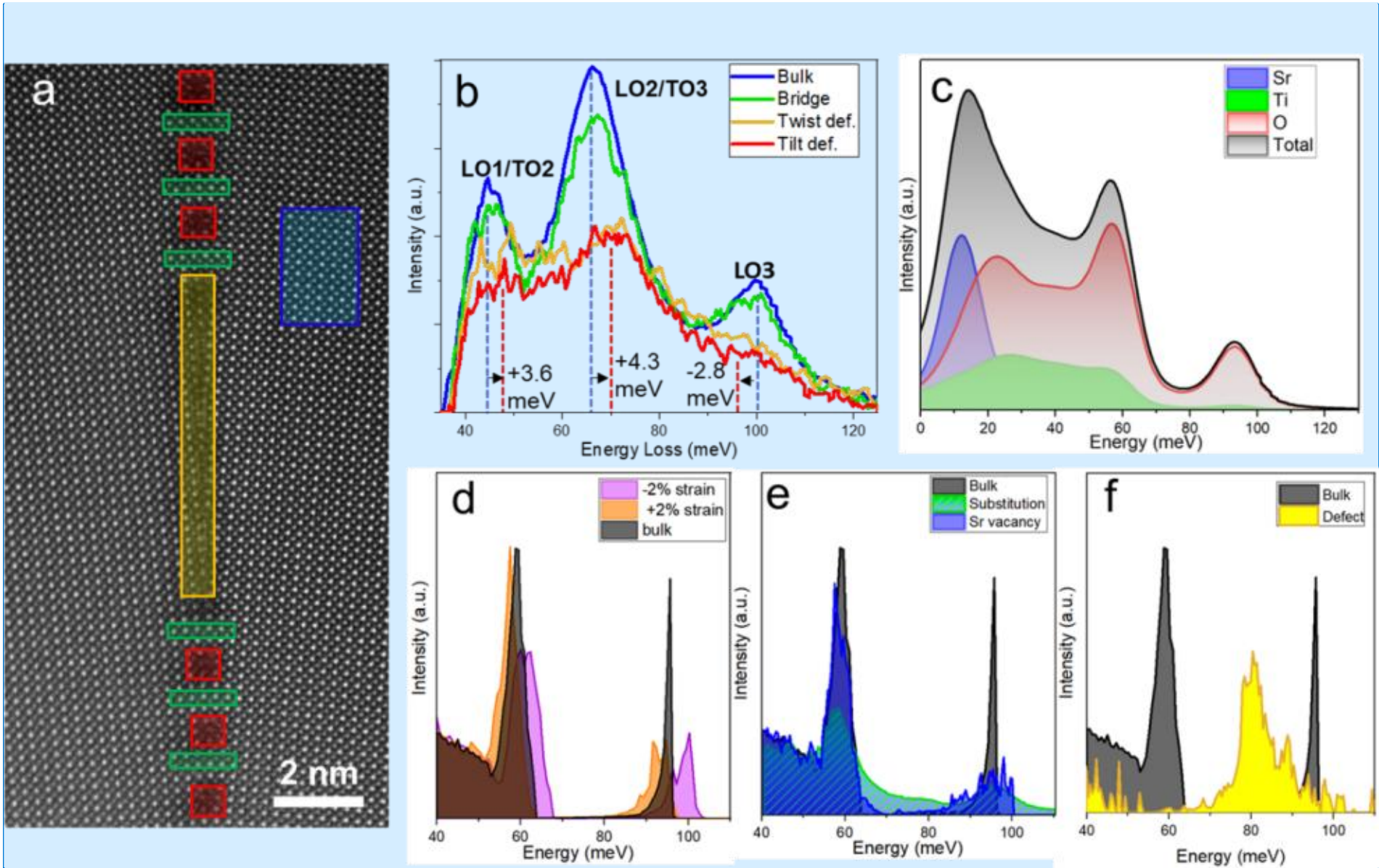

**Figure 2. Vibrational EELS and DFT calculation of site-specific PDOS.** (a) HAADF with bulk, tilt-induced core, twist-induced core, and bridge structures marked. (b) A comparison of ZLP normalized experimental spectra for bulk, tilt-induced core, twist-induced core, and bridge structures. (c) DFT calculated phonon DOS with elemental contribution convolved with probe size and compared with experimental bulk signal (d) DFT calculated PDOS of $E_{yy}$ strained bulk. (e) DFT calculated PDOS of STO with Sr vacancy and Ti for Sr ($Ti_{Sr}$) substitutional defect. (f) DFT calculated PDOS at GB defect core.

The peak position and intensity in the vibrational spectra are directly correlated with the local atomic arrangement and composition and reveal information about the available phonon states and their population. The bulk optical phonon spectrum consists of 3 phonon mode groups due to the cubic symmetry: LO1/TO2 (Ti-O vibration) modes centered around 46 meV, LO2/TO3 (Ti-O and O-O vibrations) modes centered around 64 meV, and LO3 (Oxygen octahedral breathing modes) mode centered around 97 meV[39,40]. The phonon modes of the defect cores are less distinct owing to the structural disorder and loss of cubic symmetry. The LO1/TO2, LO2/TO3, and LO3 modes undergo a reduction in intensity of 25%, 49%, and 59%, respectively, compared to those in the bulk STO. The reduction in phonon intensity suggests a decrease in occupied phonon states. Lower states at the core are expected to impede cross-GB thermal transport[3,41]. In the twist-induced core, the LO1/TO2 to LO2/TO3 peak height ratio is 0.98, in contrast with the 0.86 in tilt-induced core, implying that the stacking fault along the z-axis (Fig. 2b), in the twist-induced core is less favorable for longitudinal modes.

Additionally, the TO1/TO2 and LO2/TO3 modes experience a blueshift of 3.6 and 4.3 meV, respectively. The LO3 peak, however, exhibits a redshift of 2.8 meV. The energy shifts of the

**Commented [XW2]:** c below d, d, e, f to right.

**Commented [XW3R2]:** label areas on (a)

**Commented [XW4R2]:** label tilt twist region major phonon modes represent an overall broadening of the experimental phonon density of states shape[42,43]. This broadening of states in energy in the defect cores creates a large mismatch of states between the bulk and defect regions further impeding phonon transport. Despite stacking fault in the z-axis, the twist-induced core exhibits similar vibrational signature as the tilt-induced core with a slightly enhanced TO1/TO2 peak. For the 22° defect phonon spectrum we see an even less pronounced LO2/TO3 and LO3 peak. By comparing bulk, bridge, 10° defects, and 22° defect we can see a pattern of gradual transition from most bulk-resembling to most defect resembling of the phonon spectra.

For comparison, Fig. 2c shows the DFT phonon density of states (PDOS) of bulk STO. The PDOS of STO with strain, vacancies, and substitutional defects are separately computed in order to explore their independent contributions to the GB vibrational properties. Compressive and tensile strains only cause slight blueshifts and redshifts of phonon modes respectively (Fig. 2d), while vacancies result in small changes in the PDOS peak intensity and emergence of only a few modes in the phonon bandgap (Fig 2.e). Most notably, our results in Fig. 2e illustrate that elemental substitution of Sr by Ti causes the greatest change to the PDOS with the emergence of additional modes in the phonon bandgap and the spreading out of phonon modes in energy. These results are consistent with the Ti $L_3$ $t_{2g}$ width which shows that the tilt and twist induced cores have a similar degree of Ti valence reduction, indicating a similar degree of Ti substitution of Sr. This similarity explains their resemblance in vibrational signals despite their different atomic arrangement such as the presence of a stacking fault in the twist-induced core. By isolating structural and chemical components of GBs, our results generalize to a wide variety of tilt and twist combinations. Fig. 2f shows the calculated PDOS of a full GB defect core. The defect core modes show the LO2/TO3 peak has blueshifted and the LO3 peak has redshifted to the extent that they combined into one single peak at 80 meV. This is similar to experimental observation of peak shifts. Compared to simulations of bulk structures with elemental substitution and vacancies, we observe emergent phonon behavior, such as the redshift of the LO3 peak near 100 meV.

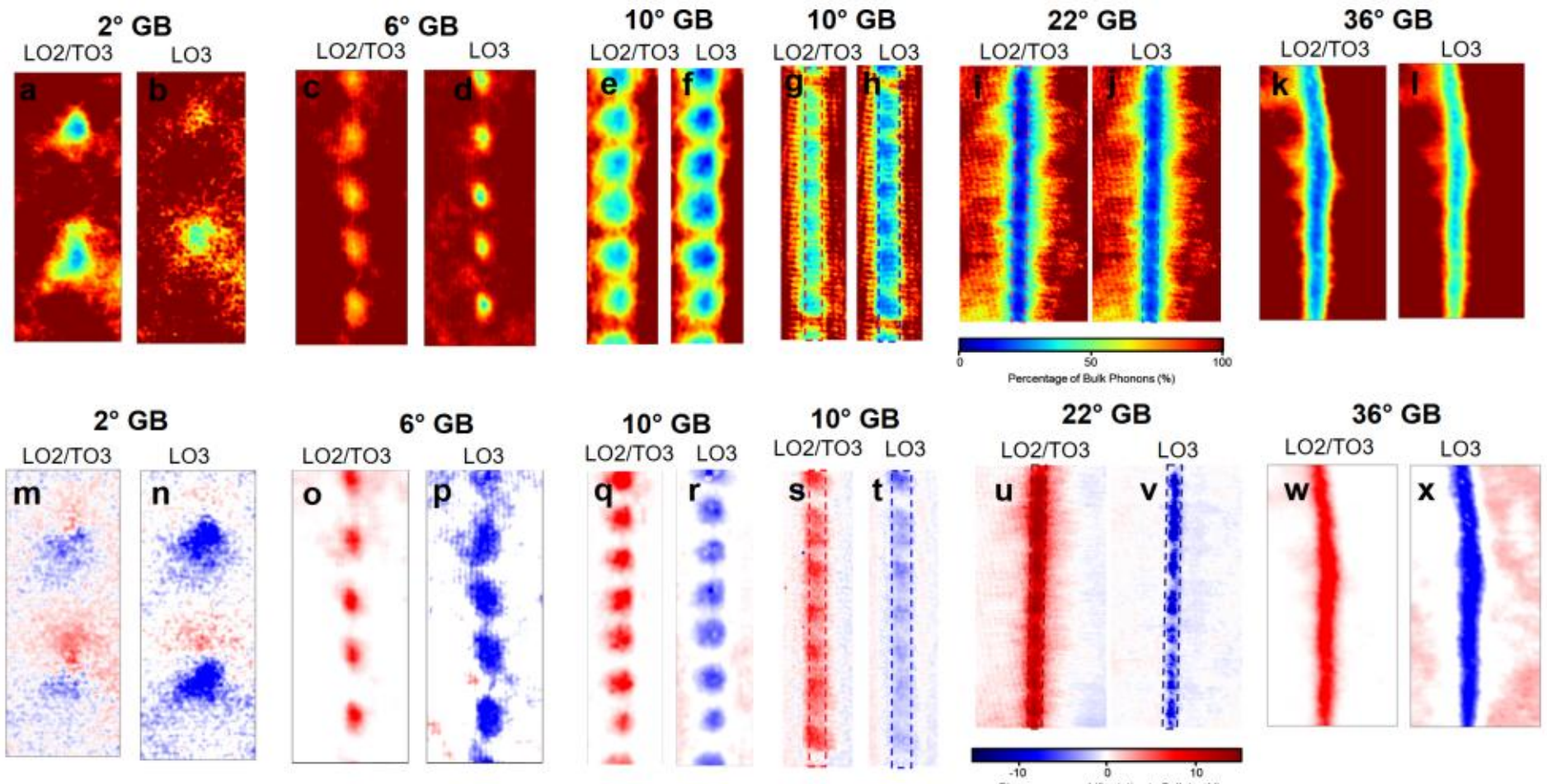

**Fig. 3 | Mapping of LO2/TO3 and LO3 phonons around 10° and 22° GB defects.** The width of all mappings except (e), (f), (g), (h), (q), (r), (s), and (t) are 8nm. (a), (c), (e), (g), (i), (k)Mapping of percentage of LO2/TO3 phonon population relative to bulk of 2°, 6°, 10°, 10°(0.34°), 22°, and 36°GB respectively. (b), (d), (f), (h), (j), (i) Mapping of percentage of LO3 phonon population relative to bulk of 2°, 6°, 10°,10°(0.34°), 22°, and 36° GB respectively. (m), (o), (q), (s), (u), (w) Mapping of percentage of LO2/TO3 phonon peak shift relative to bulk of 2°, 6°, 10°, 10°(0.34°) , 22°, and 36°GB respectively. (n), (p), (r), (t), (v), (x) Mapping of percentage of LO3 phonon peak shift relative to bulk of 2°, 6°, 10°, 10°(0.34°), 22°, 36° GB respectively.

A spatial mapping of the major longitudinal and transverse phonon modes reveals tilt and twist induced modulations in the relative phonon peak areas (Fig. 3a-l) which are proportional to the number of occupied phonon states. This can be seen from the scattering cross-section for vibrational EELS which has 3 main terms: A phonon occupancy term *n+1*, a selection rule-like term dependent on the effective nuclear charge, and finally the phonon density of states[3]. In the tilt defect cores of the 2°, 6°, and 10° GB, occupied phonon states dramatically decrease with respect to the bulk (Fig. 3a-h and Extended Data Fig. 7, Ext. Data Fig. 8) owing to a lack of atomic species and increased disorder. In the bulk-like bridge structures, however, the phonon population is only slightly reduced, suggesting pathways for phonon transport. Although the phonon population in the twist defect structure is similar to that of tilt defect cores, its continuous nature suggests greater phonon impedance. The 22° and the 36° belong to the regime of high angle GB. The tilt defect cores in the 22 GB exhibit a relative phonon population much less than that of the 10 GB tilt defect cores, demonstrating the coarse tuning nature of varying the tilt angle. The continuous nature of the 22° GB lends itself as a more efficient resistor of thermal transport than the 10° GB tilt-induced cores with bridge structures. Although both the twist-induced defect in the 10° GB and the 22° GB are continuous, the more disordered crystal [35,41]structure in the 22° GB, contributes to a greater lowering of its phonon population. This again supports the idea that the twist-induced core acts as a fine-tuning structure that is similar to the tilt-induced core in its vibrational character but is disordered enough to inhibit more phonons than the tilt-induced cores. The 36° GB is similar to the 22° in that the defect is continuous. The difference is that 36° has higher level of coherency due to high density of coincidence lattice sites (CLS) as a Σ5 GB. In nanoscale phonons the continuous nature of defect phonons is repeated at the 36° GB. However, the effect of periodic CLS is not visible in the phonon mapping as no periodic change in phonon population in correspondence with CLS periodicity is observed. This atomic-scale mapping of phonon modes demonstrates that by manipulating grain tilt and twist, the spatial periodicity and dimensions of defect cores can be controllably tailored. This enables deterministic modulation of nanoscale phonon states at GB interfaces. In addition to limiting phonon transport via reduction of states, phonon energy mismatch also greatly impacts phonon propagation[1]. Patterns of phonon energy shift generally resembles that of phonon population in that they are very localized to defects. We also see a blueshift of LO2/TO3 mode and a redshift of LO3 mode for all GBs. Tracking the energy shift of the LO2/TO3 and LO3 vibrations gives us a measure of how dispersed the phonon modes are and their mismatch between bulk and GB modes. Phonons in both the tilt and twist induced cores of the 10° GB and phonons in the 22° GB exhibit a blueshift of the LO2/TO3 modes and a redshift of the LO3 mode, while little to no energy shift is found in bridges compared to the bulk (Figs. 3q-v). The energy shifts in the 22 GB are more drastic, suggesting that controlling the tilt angle coarsely tunes the energy shifts while varying the twist angle has a much finer influence.

Unlike the phonon population which has a spatial localization with a full-width at half maximum (FWHM) of 1.25nm, the phonon energy shift is highly localized to the cores with a full-width at half maximum (FWHM) of 0.67 nm (Extended Data Fig. 8). The spatial distribution of the energy shift is consistent with that of the HAADF intensity of 0.64 nm, indicating that phonon mode energy is strictly localized to the structural disorder at the core. This suggests that the phonon bands, reflected in the PDOS shape, are highly localized, whereas the local phonon population is delocalized beyond the cores. This is likely due to boundary conditions at core edges restricting large-wavelength phonons or a restriction in the propagation direction of phonons.

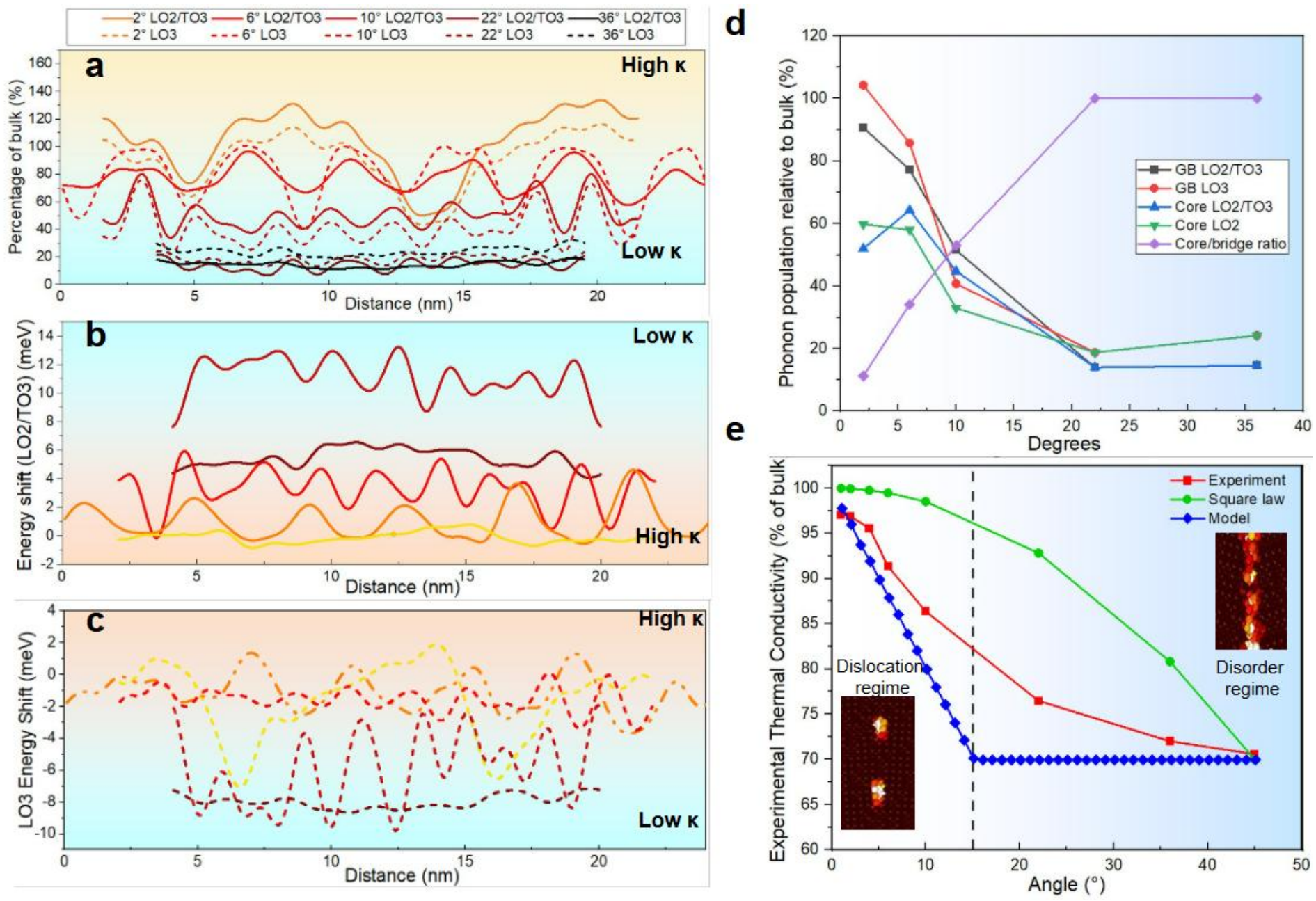

**Fig. 4 | Quantification of GB defect and Bulk phonons.** (a) Line profiles of LO2/TO3, and LO3 phonon populations of all the GBs. (b) Line profiles of LO2/TO3 phonon relative peak shift of all the GBs. Line profiles of LO3 phonon relative peak shifts of all the GBs. (d) GB phonon population averages, core phonon population, and core density plotted against angles. (e) The TTG measurements, the classical square law, and our proposed theoretical GB modelled thermal conductivity in variation with GB tilt angles..

Figure 4 presents a quantitative description of how much phonons are reduced compared to bulk for each GB and correlates this nanoscale evidence with macroscopic thermal transport measurements. In the small angle tilt GBs (2°, 6°, and 10°), the phonon populations dip at each distinct core along the GB. The 2° GB has a defect periodicity of roughly 8nm. The 6° GB has similar magnitude of phonon population decrease at the cores, but its smaller core periodicity means this decrease happens more often along the GB. The 10° GB is special in that it shows the coexistence of two types of defects. In the tilt defect of the 10° GB, occupied phonon states dramatically decrease to an average of 54% of the bulk (Fig. 4a and Extended Data Fig. 7) while the twist defects exhibit an average of 42%. The 22° GB is where the defect cores merge and covers the entire GB. In the 22° GB, phonons at the tilt defect are down to an average of 16% of their bulk quantity. Comparing GB phonons directly, tilt defect phonons in the 10° GB are about 3 times more populous than those at the 22 GB tilt defect, implying that the more disordered and continuous 22 GB is more efficient at impeding phonons. Fig.4 a shows that there are corresponding alternating regions of low and high peak areas in the twist-induced core, with the highest peak still far lower than that of bridges between tilt-induced cores. Although the twist defect at the 10° GB and the 22° tilt defect are both continuous, the twist defect at the 10° GB has a nearly identical vibrational signature and defect core phonon population as its 10° tilt defect counterpart. In the 36° GB we see a similar phenomenon as in the 22° GB where the defect cores are merged. The 36° GB has high density of coincidence lattice sites (CLS), being a Σ5 GB. Despite the high density of CLS, the 36° GB shows a low phonon population as in the 22° GB.

Fig 4.d summarizes nanoscale phonon findings of all GBs. In summary, core phonon population decreases and the ratio of defect cores occupying the GB increases as the tilt angle increases. Both of these trends, however, would gradually saturate as the tilt angle surpasses the critical angle where the defect covers the entire GB. The pattern of phonon peak shifts can be summarized as it shows similar patterns with phonon populations in that: first the peak shift occurs at the defect cores for the low angles (2°, 6°, and 10°) and covers the entire GB at high angles (22°, and 36°). The peak shifts are similar in that the LO2/TO3 mode shows blueshift while LO3 mode shows redshift at defects at all GBs.

To quantify the impact of the observed microstructural differences in the GBs on macroscopic thermal transport, we used transient thermal grating (TTG) spectroscopy to measure the in-plane thermal conductivity across GBs with different tilt angles (details in Supplementary Information). Figure 4e shows the raw traces of the TTG measurement. When the TTG measurement was conducted far away from the GBs, the measured effective thermal conductivity in both samples was around 11.3 $Wm^{-1}K^{-1}$. When on the GBs, different GB show discernible thermal conductivity differences. We can see that thermal conductivity decreases with tilt angle, but the manner of decrease is different from classical model where thermal reflectance is proportional to the square of the tilt angle (green plot). Instead, it first decrease rapidly, but seem to saturate at higher angles. This finding is in congruent with our nanoscale phonon measurements. Where defect core density saturates in the high tilt angle regime. To better summarize our findings, we made a model that encapsulates GB thermal behavior. In short, GB thermal conductivity has the following relationship with the tilt angle:

$$K_{tot} = \left(1 - \frac{2\sin\left(\frac{\theta}{2}\right)}{a} K_b\right) + \frac{2\sin\left(\frac{\theta}{2}\right)}{a} K_d$$

Where $k_{tot}$ is total thermal conductivity, $k_b$ is bulk thermal conductivity, $k_d$ is dislocation thermal conductivity, a is lattice constant, and θ is tilt angle. This model is plotted in Fig 4. e and it somewhat captures the relationship between tilt angle and GB thermal conductivity. In comparison, the reduction in the thermal conductivity by the 22° GB (21.8%) is appreciably higher than that by the 10° GB (15.6%). This result indicates that the observed microstructural variation in the GBs in this work significantly influences the macroscopic thermal transport property. The short periodicity (66nm) of the twist-induced defect means that its effect on thermal conductivity cannot be directly measured using TTG, which has spatial resolution at the micrometer level. The different effect of thermal conductivity of the tilt and twist induced cores was simulated with MD (Extended Data Fig.10), where the Kapitza conductance is 1.18 GW $m^{-2} K^{-1}$ for the tilt-induced core and 0.64 GW $m^{-2} K^{-1}$ for the twist-induced one. Nanoscale phonon mapping and macroscale thermal conductivity measurements and simulation results strongly suggest that tilt and twist angle variations have a coarse and fine-tuning effect that can be exploited for highly precise modulation of thermal conductivity.

In summary, establishing a direct relationship between GB structure and vibrational properties is crucial for engineering GBs for optimizing thermal conductivity for applications. In this study, we have revealed a twist and tilt manipulation on a single GB can impact crystalline thermal conductivity and revealed its underlying atomic-scale mechanism in terms of local GB phonon populations and modes. We categorized symmetric tilt GB as two types: low angle and high angle. Low angle tilt GBs create edge dislocation cores at the GB with defect core spacing roughly inversely proportional to the tilt angle. Reducing inter-dislocation spacing drastically decreases phonon transport and thermal conductivity. In the high angle regime, dislocation occupation at the GB is saturated. Further change in tilt angle in this regime leads to little change in thermal conductivity. We found that the effect of coincidence lattice sites has very limited, if detectable changes in phonon signature and thermal conductivity. Changes in tilt angle create a global change in the phonon spatial distribution at the GB and have a drastic effect on macroscopic thermal conductivity. Twist, however, creates twist-induced defect region coexisting and alternating with the tilt dependent tilt-induced defect region. The unique phonon distribution and simulated thermal conductivity at this twist-induced region produce the effect of fine, spatially defined changes on the GB thermal conductivity. This demonstrated tunability is of great interest to the field of nano-engineering as it allows for the continuous adjustment of cross-grain boundary phonon transport properties at nanoscale precision confined to a 2-dimensional plane while preserving other intrinsic electrical and mechanical properties of the material elsewhere. This implication is paramount in offering an additional dimension in controlling the property and functionality of polycrystalline and composite materials.

## Methods

### Bicrystal Fabrication and STEM sample preparation

**Commented [XW5]:** add hand polishing

The STO bicrystal was prepared through high temperature annealing of two separate pieces of 5° tilt mono-grain bulk STO crystal. Before annealing the surfaces of the symmetric tile crystals are polished to be atomically smooth and cleaned off contaminants using ethanol and propanol. The annealing process then began with the crystals being aligned with 0.5° accuracy in any direction. The first phase of annealing involves applying a 0.2 MPa uniaxial pressure and at 700 °C for 20 h. The second phase is to apply 1000 °C for 80 h for a larger bond area. The resultant crystal then has a [001]/(100) 10° tilt. The dimension of the bulk bicrystal sample is 2×15×1 mm with the GB being parallel to the 2×1 mm face, splitting the 15mm edge in half. The 0.34° twist is believed to be caused by the 0.5° uncertainty in alignment around the (010) axis. The 22° bicrystal is fabricated using the same method but with a different tilt angle.

TEM samples are then prepared using focused ion beam (FIB) in-situ lift out technique. 2 µm tall 1.5 µm wide and 15 µm long platinum layers are deposited across the GBs to protect the to-be-lift-out lamellae containing the GBs. two 15×10 µm trenches are made along the GB and on either side of the platinum layer. The trenches are 6 µm deep. A L-shaped cut is then performed on two of the 3 edges on the lamellae attached to the bulk. An Omni-probe needle is then attached to the detached side of the lamellae and welded using platinum deposition. The remaining attached side is then cut off and the lamellae is then lifted. The lamellae on the needle is then welded on a standard TEM copper grid and further thinned using FIB until the thinnest part reaches a thickness of 30 nm. The copper grid is then loaded into Fishone Nanomill, which uses an Ar ion beam to clean off the amorphous layer caused by the FIB preparation. The milling time is set to 10 min on each side of the lamellae under 700 eV voltage each session. Liquid nitrogen was applied during milling to avoid structural damage during the fine milling.

**Calculation of GB twist and tilt angles and their corresponding defect periodicities**

At the GB, the structures are discrete, meaning that it can be either structure A or structure B and cannot be a structure intermediate between the two. In other words, the planes of the two crystals are either exact aligned as in structure A or exactly in between each other as in structure B. This suggests the atoms are strained from their positions when the crystals are purely twisted and the GB configuration are not allowed to relax. Using a simple trigonometric calculation, we find the misalignment distance (distance of actual atomic location to supposed atomic location of undistorted twisted grains) is $\Delta z = \pm 164.5$ pm on either side of the twist-induced defect core. This indicates the maximum misalignment of the (001) planes until the twist-induced core mechanically relaxes to a stacking fault. We can thus formulate a crude relationship between the length of the twist-induced core $l$ and the twist angle $\beta$ as $l = (d/2 - \Delta z)*\cos(\alpha)$.

**HAADF Imaging, EDS, and core-loss EELS data collection**

High magnification atomic resolution HAADF imaging was performed on JEOL 300 GrandARM microscope in STEM mode under 300 keV electron beam and 33 mrad convergence angle. EDS and coreloss EELS data are collected under the same beam condition as HAADF imaging. Collected core-loss EELS data are background subtracted and peak separated to determine the peak position, onset, width, height, and intensity. The Ti-L edge is observed in 4 distinct peaks for bulk STO. The energy range is 450-480 eV. High energy resolution is not required for core-loss mapping. The O-K edge is in the energy range of 530-550 eV and shows 3 major peaks. The step size is set to one angstrom for core-loss mapping. The map area is set to include both tilt and twist induced defect cores in a symmetrical manner. Representative spectra of the defect cores and bulk are obtained by summing the signals of multiple corresponding pixels. This information is then mapped to reflect changes in core-loss data at the defect cores along the GB.

**Analysis of EDS mapping**

A detailed analysis of the atomic structure, provided by the strain and elemental distribution of the GB structures (Extended Data Fig 3 1-j, and Extended data Fig.4 a-j), is necessary to understand GB localized phonon modes and how they affect thermal transport. Extended Data Fig.3 shows tensile/compressive strain along the cross-GB direction by mapping unit cell distortions using affine transforms. Strain maps in Extended Data Fig.3 show that in general, bridge structures have minimal strain, while tilt and twist induced cores have high strain. The EDS mappings in Figs. 4reveal a strontium deficiency and titanium surplus in both tilt and twist induced cores while the percentage composition of oxygen is roughly uniform (Fig. 4d). Note that the strontium composition inversely correlates to titanium composition (Fig. 4e). In both tilt and twist-induced defect cores, strontium exhibits a 5 percent deficiency while Ti has a 5 percent surplus, indicating noticeable element substitution in GB regions. It is surprising that the twist-induced core contains similar periodic regions of Sr vacancies and Ti surplus as the tilt-induced defect core, even though it seemingly has a more complex atomic arrangement. The compositional similarity between the twist-generated twist-induced core and the tilt-generated core for the 10° tilt and 0.34° twist suggests that varying the twist angle finely changes the chemistry of the GB. Varying the tilt angle to 22° induces a Sr surplus and Ti deficiency at the 22° GB.

**Analysis of Core loss fine structure**

A deeper understanding of the GB structure is obtained upon probing the Ti and O electron energy loss near edge structure (ELNES) (Extended Data Fig. 5k-q). In bulk STO, Ti is in the $4^+$ valence state and has six-fold bonding symmetry[42–44], which gives rise to 4 distinct peaks ($t_{2g}$ and $e_g$ for $L_2$ and $L_3$). At the defect cores, the Ti-L edge is very sensitive to change in bonding coordination caused by missing neighbors and change in oxidation state caused by Sr vacancies and Ti surplus (Extended Data Fig. 5k-m). Bonding coordination changes cause symmetry breaking which is reflected in an overall broadening of the $e_g$ peak at the grain boundary (Extended Data Fig. 5h). In addition, Ti substitutions in Sr sites (Extended Data Fig. 4b-c) in GB cores cause reduction from $Ti^{4+}$ to $Ti^{3+}$. This results in a broadening of both $L_3$ $t_{2g}$ and $e_g$ peaks and a redshift of the $e_g$ peak (Extended Data Fig. 5h). As the broadening of the $L_3$ $t_{2g}$ peak can only be a result of the existence of $Ti^{+3}$ species, its mapping shows the locations of $Ti^{+3}$ species, which are highly localized in the GB cores (Extended Data Fig. 5). The peak width seems to be inversely correlated with low HAADF intensity (defect cores) (Extended Data Fig. 5d). It is worth noting that the peak characteristics at the bridge regions closely resemble that of the bulk, indicating a structural and compositional similarity. Furthermore, The $L_3$ $t_{2g}$ width of the twist-induced core resembles that of the tilt-induced core, whereas the bridge-like core-loss signature is not present in structure B due to its [001] stacking fault-like structure. The ELNES mapping of the 22° GB is shown in figure 2. n-p. The spatial distribution of defect core-loss signal is similar to twist-induced defect with the twist-induced defect in that there's no inter-defect bridge structure. For clarity, the representative coreloss structure of all structures is shown in Extended Data Fig 5.h.

**Strain Mapping**

To extract the cell deformation information, we first extract the atom position using the software Atomap[45]. Each atom cell block shape can then be reconstructed by matching the typical shape of a Perovskite Oxides, where the structure contains four atoms on the edge and one atom in the center. For each cell block, four outer atoms are mapped back to a reference block, where the reference is typically a standard undeformed cell structure or an average cell shape. The affine transformation matrix can be computed using the estimateAffine2D function implemented in opencv package, and the optimal 2D affine transformation matrix H is computed using random sample consensus method such that [x y] = [a11 a12; a21 a22] * [X; Y] + [b1; b2].

Strain mapping was generated based on HAADF images by fitting single atom positions and using affine transform to obtain the transformation coefficients (Extended Data Fig. 2) and separately using geometric phase analysis (Extended Data Fig. 3). Strain mapping using affine transforms shows better performance.

**Molecular Dynamics**

MD simulations are carried out to better elucidate core formation mechanisms. Empirical MD potentials are applied to reduce computational cost and allow large scale simulation of the GB. LAMMPS is chosen to be the simulation software. A simulation cell of size 66.5×28.5 × 2.4 $nm^3$ is used with the GB along the 66.5nm side and bisecting the cell. The two crystals are set to have a [001]/(100) 10° tilt with a variety of [010]/(100) twist angles including 0, 0.34, 0.68, 1.02, and 1.36° twist angles. Apart from the 10° tilt, a 22° a [001]/(100) tilt is also built. To avoid atoms overlapping a 2 Å cutoff distance is set to remove atoms too close to each other. Energy minimization and position relaxation is then carried out for all the atoms to reach an ultimately stable structure[46,47].

**DFT Simulation**

DFT calculations of bulk phonon DOS are performed using undistorted unit cells of STO. The structure is first relaxed and then force constants are obtained based on the relaxed structure. The obtained force constants then undergo matrix transformation to generate the phonon DOS and dispersion curve. The phonon DOS is then convolved with a 10meV energy resolution ZLP to be better compared to the experimentally measured data. The PDOS involving strain, vacancy, and elemental substitution are performed over 4x4x4 unit cells using a similar procedure[48–51]. The structure of a 22° GB is also built separately from that of the MD simulations. The structure is much smaller, containing 246 atoms within an unit cell of 0.39×2.0×3.95nm unit cell. The cell is then structurally relaxed and phonon DOS calculated. The 10° cell simulation was not included because more atoms are necessary to capture a single period of tilt-induced cores, which is unachievable due to high computational cost from DFT.

**Vibrational EELS data collection and analysis**

Nion UltraSTEM 200 microscope with "HERMES" system was used to collect vibrational data, operating at 60 kV. In addition, 33 mrad convergence semi angle is used to enhance spatial resolution. The post specimen beam is then deflected by 60 mrad by a combination of post specimen lenses so that the direct beam is just outside of the EELS detector to avoid delocalized signal from phonon polariton coupling inherent in polar materials[52]. The resultant FWHM of the zero-loss peak (ZLP) is 11-14 meV, enough to resolve phonon peaks of STO. The beam is then raster scanned through the area of interest, and dwell time of each pixel is set to 1 second. The size of the pixel step size is set to 1 Å. The EELS spectrum of each pixel is then filtered through principal component analysis to reduce noise level using first 3-5 components, and then background subtracted using a combination of Gaussian and Lorentzian fitted ZLP. The subtracted signal is then normalized against the ZLP, and then deconvolved using Pseudo-Voigt deconvolution into 3 peaks, with information about peak height, width, area, and position calculated. This information is then mapped and gaussian blurring is applied to the pixelated maps for better aesthetics.

**Thermal transient grating (TTG)**

Thermal transient grating method which is sensitive to across-grain boundary thermal conductivity is used to measure the thermal conductivity at the grain boundary and bulk of the 10°tilt and 22°tilt bicrystal samples. TTG uses an optical pump probe method to detect thermal conductivity. Two pump laser beams are used forming a 1D periodic spatial grating along the in-plane (across GB) direction. The pump beams trigger a material response through absorption and photo-electron interaction. The excited electrons then transfer thermal energy to nearby electrons and atoms through electron-electron interaction and electron-phonon interaction. The grating thus produces periodic thermal excitation on the material. The response is then detected by the diffraction of a probe beam. The decay of material excitation through time is recorded and fitted to obtain thermal conductivity. Prior to measurement, 65nm aluminum is deposited on the surface for better absorption. The probe size is 200μm and the power for the pump is 50mW and for the probe is 17 mW[52].

**Author Contributions**

X.Q.P. conceived and directed this project. X.W.W. and C.A.G. designed the experiments. X.W.W., C.A., and M.X. prepared the cross-sectional samples. X.W.W. carried out the EELS experiments with the help of C.A.G, X.X.Y, and T.A. Vibrational EELS data was analyzed by X.W.W using custom code developed by C.A.G. C.A. and C.J.D. conducted imaging, core-loss EELS, and EDS experiments and analyzed the data with the help of C.A.G. X.W.W, C.A.G, R.Q.Y carried out TTG experiments on the GBs with the help of Y.J.Q and supervision of B.L.L.. W.J.Z conducted MD simulation on thermal conductivity. X.B, and P.H.C interpreted the observed vibrational, core-loss, and structural phenomena. B.X. carried out MD simulations on all the GB structures. W.T.P. carried out DFT phonon calculations under the direction of S.P.O. Y.F.H. wrote custom code to map unit cell distortions and obtain strain measurements and F.G. conducted GPA mapping. The manuscript was prepared by X.W.W., C.A.G., X.X.Y., R.Q.W. B.L.L, S.P.O., P.H.C., X.Q.P. with input from all co-authors.

**Acknowledgement**

All experiments are conducted using the facilities in the Irvine Materials Research Institute (IMRI). The MD simulations are conducted using UC Irvine Research Cyberinfrastructure Center (RCIC). DFT simulations are conducted using San Diego Supercomputing Center (SDSC). This work was primarily supported by the National Science Foundation Materials Research Science and Engineering Center (NSF MRSEC) through the UC Irvine Center for Complex and Active Materials under the Grant No. DMR-2011967 and partially supported by the (Department of Energy (DOE), Office of Basic Energy Sciences, Division of Materials Sciences and Engineering under Grant No. DE-SC0014430. Acknowledgement to Prof. Yu Pu for supplying the 2° and, 6° samples. Acknowledgement to Dr. Haiping Sun for many of the TEM sample preparations.

**Commented [XW6]:** Acknowledge Yu Pu and Haiping.

**References**

1. Gadre, C. A. *et al.* Nanoscale imaging of phonon dynamics by electron microscopy. *Nature* **606**, 292–297 (2022).

2. Yan, X. *et al.* Single-defect phonons imaged by electron microscopy. *Nature* **589**, 65–69 (2021).

3. Cahill, D. G. *et al.* Nanoscale thermal transport. II. 2003–2012. **011305**, (2019).

4. Kim, S. Il *et al.* Dense dislocation arrays embedded in grain boundaries for high-performance bulk thermoelectrics. *Science (80-. ).* **348**, 109–114 (2015).

5. Wang, Z., Alaniz, J. E., Jang, W., Garay, J. E. & Dames, C. Thermal conductivity of nanocrystalline silicon: Importance of grain size and frequency-dependent mean free paths. *Nano Lett.* **11**, 2206–2213 (2011).

6. Krivanek, O. L. *et al.* Vibrational spectroscopy in the electron microscope. *Nature* **514**, 209–212 (2014).

7. Hoglund, E. R. *et al.* Emergent interface vibrational structure of oxide superlattices. *Nature* **601**, 556–561 (2022).

8. Hage, F. S. *et al.* Nanoscale momentum-resolved vibrational spectroscopy. *Sci. Adv.* **4**, eaar7495 (2018).

9. Lu, Q. *et al.* Bi-directional tuning of thermal transport in SrCoOx with electrochemically induced phase transitions. *Nat. Mater.* **19**, 655–662 (2020).

10. Tomko, J. A. *et al.* Tunable thermal transport and reversible thermal conductivity switching in topologically networked bio-inspired materials. *Nat. Nanotechnol.* **13**, 959–964 (2018).

11. Pernot, G. *et al.* Precise control of thermal conductivity at the nanoscale through individual phonon-scattering barriers. *Nat. Mater.* **9**, 491–495 (2010).

12. Zhu, G. *et al.* Tuning thermal conductivity in molybdenum disulfide by electrochemical intercalation. *Nat. Commun.* **7**, 1–9 (2016).

13. Du, T. *et al.* Wide range continuously tunable and fast thermal switching based on compressible graphene composite foams. *Nat. Commun.* **12**, 1–10 (2021).

14. Tomko, J. A. *et al.* Tunable thermal transport and reversible thermal conductivity switching in topologically networked bio-inspired materials. *Nat. Nanotechnol.* **13**, 959–964 (2018).

15. Pop, E., Sinha, S. & Goodson, K. E. Heat generation and transport in nanometer-scale transistors. *Proc. IEEE* **94**, 1587–1601 (2006).

16. Seijas-Bellido, J. A., Rurali, R., Íñiguez, J., Colombo, L. & Melis, C. Strain engineering of ZnO thermal conductivity. *Phys. Rev. Mater.* **3**, 1–8 (2019).

17. Liu, C., Lu, P., Gu, Z., Yang, J. & Chen, Y. Bidirectional Tuning of Thermal Conductivity in Ferroelectric Materials Using E-Controlled Hysteresis Characteristic Property. *J. Phys. Chem. C* **124**, 26144–26152 (2020).

18. Du, T. *et al.* Wide range continuously tunable and fast thermal switching based on compressible graphene composite foams. *Nat. Commun.* **12**, 1–10 (2021).

19. Sood, A. *et al.* Direct Visualization of Thermal Conductivity Suppression Due to Enhanced Phonon Scattering Near Individual Grain Boundaries. *Nano Lett.* **18**, 3466–3472 (2018).

20. Hu, L., Zhu, T., Liu, X. & Zhao, X. Point defect engineering of high-performance bismuth-telluride-based thermoelectric materials. *Adv. Funct. Mater.* **24**, 5211–5218 (2014).

21. Gao, P. *et al.* Atomic-Scale Measurement of Flexoelectric Polarization at SrTiO3 Dislocations. *Phys. Rev. Lett.* **120**, 1–6 (2018).

22. Gao, P. *et al.* Atomic-scale structure relaxation, chemistry and charge distribution of dislocation cores in SrTiO3. *Ultramicroscopy* **184**, 217–224 (2018).

23. Lewin, M. *et al.* Nanospectroscopy of Infrared Phonon Resonance Enables Local Quantification of Electronic Properties in Doped SrTiO3 Ceramics. *Adv. Funct. Mater.* **28**, (2018).

24. Vikrant, K. S. N., Chueh, W. C. & García, R. E. Charged interfaces: Electrochemical and mechanical effects. *Energy Environ. Sci.* **11**, 1993–2000 (2018).

25. Liu, X. *et al.* Local electronic structure variation resulting in Li 'filament' formation within solid electrolytes. *Nat. Mater.* **20**, 1485–1490 (2021).

26. Choi, S. Y. *et al.* Assessment of Strain-Generated Oxygen Vacancies Using SrTiO3 Bicrystals. *Nano Lett.* **15**, 4129–4134 (2015).

27. Tai, K., Lawrence, A., Harmer, M. P. & Dillon, S. J. Misorientation dependence of Al2O3 grain boundary thermal resistance. *Appl. Phys. Lett.* **102**, 1–5 (2013).

28. Lejček, P. & Hofmann, S. Thermodynamics and structural aspects of grain boundary segregation. *Crit. Rev. Solid State Mater. Sci.* **20**, 1–85 (1995).

29. Poudel, B. *et al.* High-thermoelectric performance of nanostructured bismuth antimony telluride bulk alloys. *Science (80-. ).* **320**, 634–638 (2008).

30. Liao, K., Shibata, K. & Mizoguchi, T. Nanoscale Investigation of Local Thermal Expansion at SrTiO3 Grain Boundaries by Electron Energy Loss Spectroscopy. *Nano Lett.* **21**, 10416–10422 (2021).

31. Bhattacharya, A. *et al.* Grain boundary velocity and curvature are not correlated in Ni polycrystals. *Science (80-. ).* **374**, 189–193 (2021).

32. Liang, X. Impact of grain boundary characteristics on lattice thermal conductivity: A kinetic theory study on ZnO. *Phys. Rev. B* **95**, 1–7 (2017).

33. Haas, B. *et al.* Atomic-Resolution Mapping of Localized Phonon Modes at Grain Boundaries. *Nano Lett.* **23**, 5975–5980 (2023).

34. Yan, J. *et al.* Nanoscale Localized Phonons at Al2O3 Grain Boundaries. *Nano Lett.* **24**, 3323–3330 (2024).

35. Hickman, J. & Mishin, Y. Thermal conductivity and its relation to atomic structure for symmetrical tilt grain boundaries in silicon. *Phys. Rev. Mater.* **4**, 1–19 (2020).

36. Hoglund, E. R. *et al.* Direct Visualization of Localized Vibrations at Complex Grain Boundaries. *Adv. Mater.* **35**, 27–30 (2023).

37. Yan, X. *et al.* Unexpected Strong Thermally Induced Phonon Energy Shift for Mapping Local Temperature. *Nano Lett.* **19**, 7494–7502 (2019).

38. Hage, F. S., Radtke, G., Kepaptsoglou, D. M., Lazzeri, M. & Ramasse, Q. M. Single-atom vibrational spectroscopy in the scanning transmission electron microscope. *Science (80-. ).* **367**, 1124–1127 (2020).

39. Qi, R. *et al.* Measuring phonon dispersion at an interface. *Nature* **599**, 399–403 (2021).

40. Zeiger, P. M. & Rusz, J. Efficient and Versatile Model for Vibrational STEM-EELS. *Phys. Rev. Lett.* **124**, 25501 (2020).

41. Fujii, S., Yokoi, T., Fisher, C. A. J., Moriwake, H. & Yoshiya, M. Quantitative prediction of grain boundary thermal conductivities from local atomic environments. *Nat. Commun.* **11**, 2–4 (2020).

42. Susarla, S. *et al.* Atomic scale crystal field mapping of polar vortices in oxide superlattices. *Nat. Commun.* **12**, 1–7 (2021).

43. Crocombette, J. P. & Jollet, F. Ti 2p x-ray absorption in titanium dioxides (TiO2): The influence of the cation site environment. *Am. Lab.* **26**, 10811–10821 (1994).

44. Ohtomo, A., Muller, D. A., Grazul, J. L. & Hwang, H. Y. Artificial charge-modulation in atomic-scale perovskite titanate superlattices. *Nature* **419**, 378–380 (2002).

45. Nord, M., Vullum, P. E., MacLaren, I., Tybell, T. & Holmestad, R. Atomap: a new software tool for the automated analysis of atomic resolution images using two-dimensional Gaussian fitting. *Adv. Struct. Chem. Imaging* **3**, (2017).

46. Ramadan, A. H. H. & De Souza, R. A. Atomistic simulations of symmetrical low-angle [100] (01l) tilt boundaries in SrTiO3. *Acta Mater.* **118**, 286–295 (2016).

47. Plimpton, S. Fast Parallel Algorithms for Short-Range Molecular Dynamics for the United States Department of Energy under Contract DE.ACO4-76DPOO789. *J. Comput. Phys.* **117**, 1–19 (1995).

48. Togo, A. & Tanaka, I. First principles phonon calculations in materials science. *Scr. Mater.* **108**, 1–5 (2015).

49. Blöchl, P. E. Projector augmented-wave method. *Phys. Rev. B* **50**, 17953–17979 (1994).

50. Perdew, J. P., Burke, K. & Ernzerhof, M. Generalized gradient approximation made simple. *Phys. Rev. Lett.* **77**, 3865–3868 (1996).

51. Kresse, G. & Furthmüller, J. Efficient iterative schemes for ab initio total-energy calculations using a plane-wave basis set. *Phys. Rev. B - Condens. Matter Mater. Phys.* **54**, 11169–11186 (1996).

52. Choudhry, U. *et al.* Characterizing microscale energy transport in materials with transient grating spectroscopy. *J. Appl. Phys.* **130**, (2021).

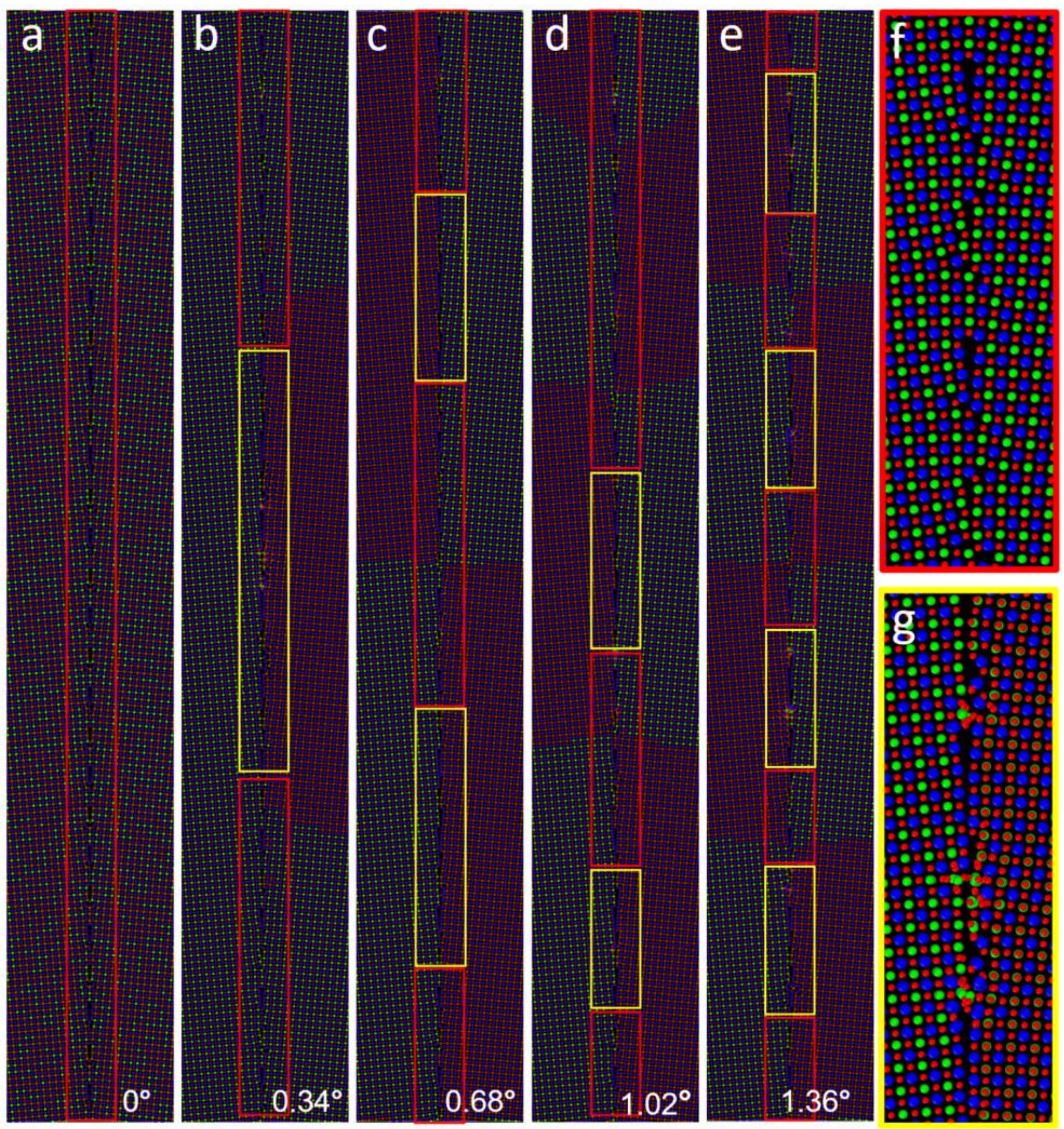

**Extended Data Fig. 1| Structural relation with twist angle.** (a) MD simulation of a twist angle of 0°. (b)MD simulation of a twist angle of 0.34°. (c) MD simulation of a twist angle of 0.68°. (d) MD simulation of a twist angle of 1.02°, one twist-induced defect on the top unexpectedly did not appear. (e) MD simulation of a twist angle of 1.36°. (f) Enlarged view of structure A (g) Enlarged view of structure B.

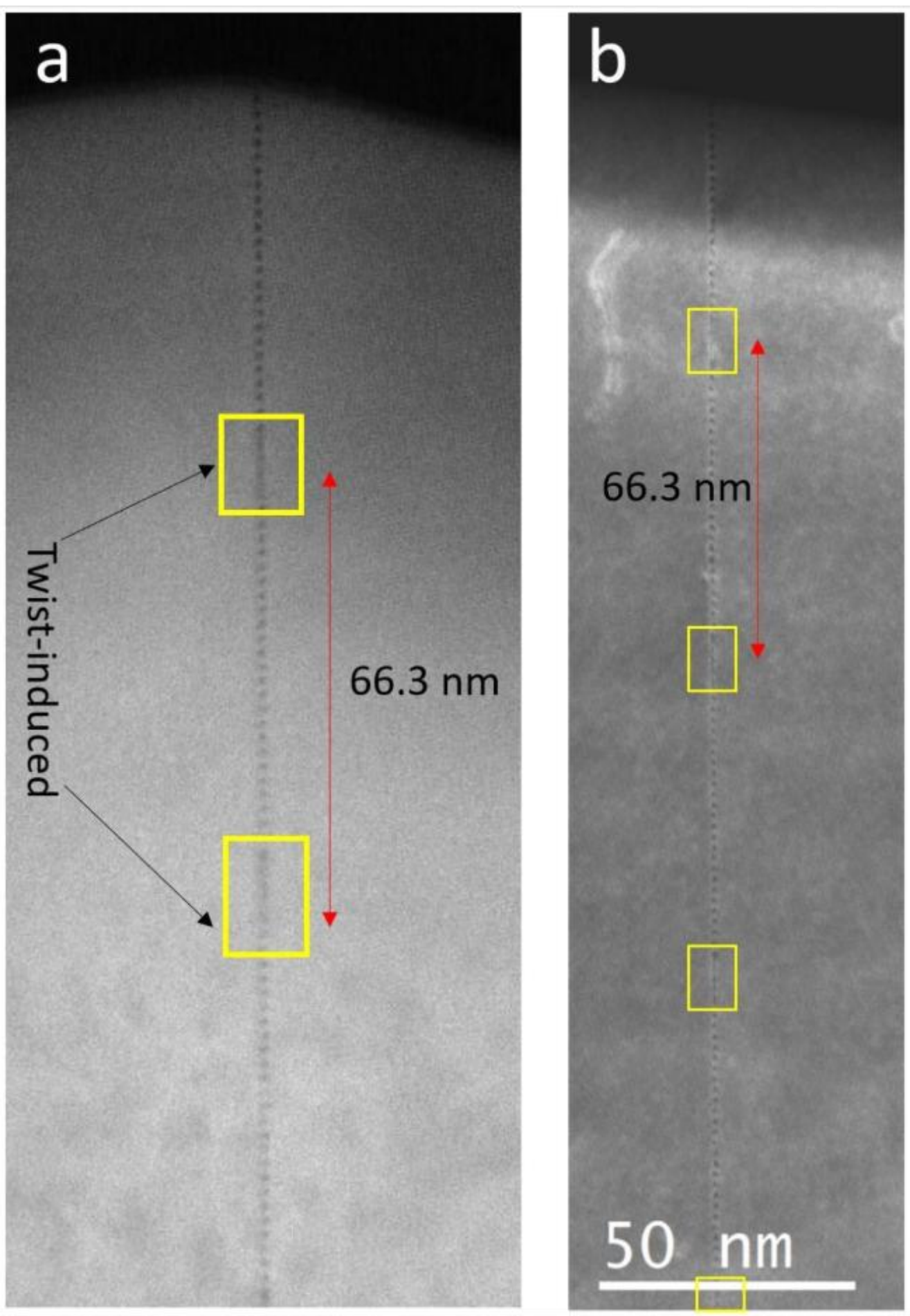

**Extended Data Fig. 2|** (a)low magnification image showing the regular periodicity of the twist induced cores. (b) even lower magnification imaged showing more than 3 equally spaced twist induced cores.

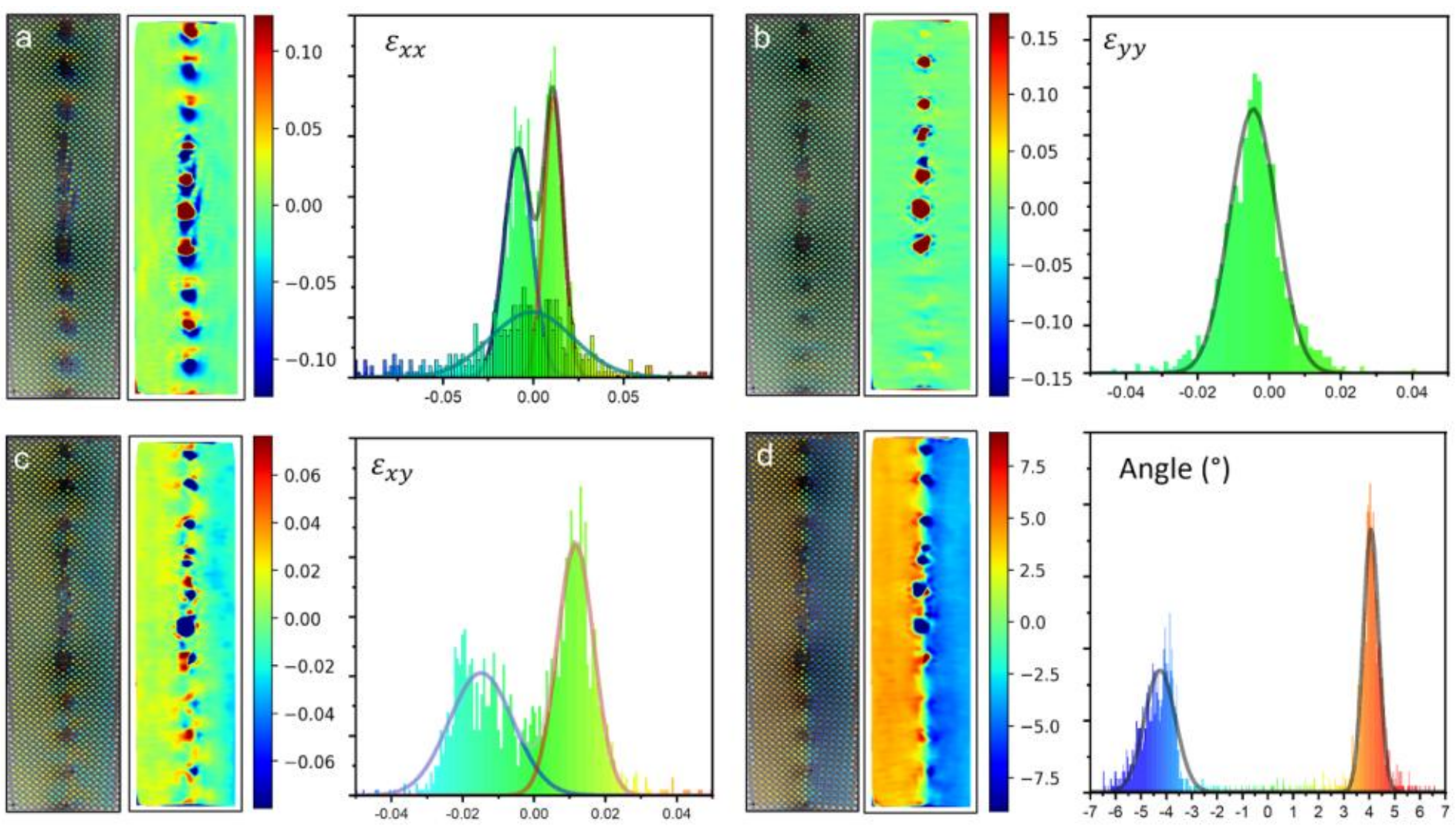

**Extended Data Fig. 3 | Strain Analysis of a 10° GB.** (a) $E_{xx}$ (b) $E_{yy}$ (c) $E_{xy}$ (d) degree of rotation, calculated at each unit cell, gaussian interpolated, and tallied according to values in bar graphs.

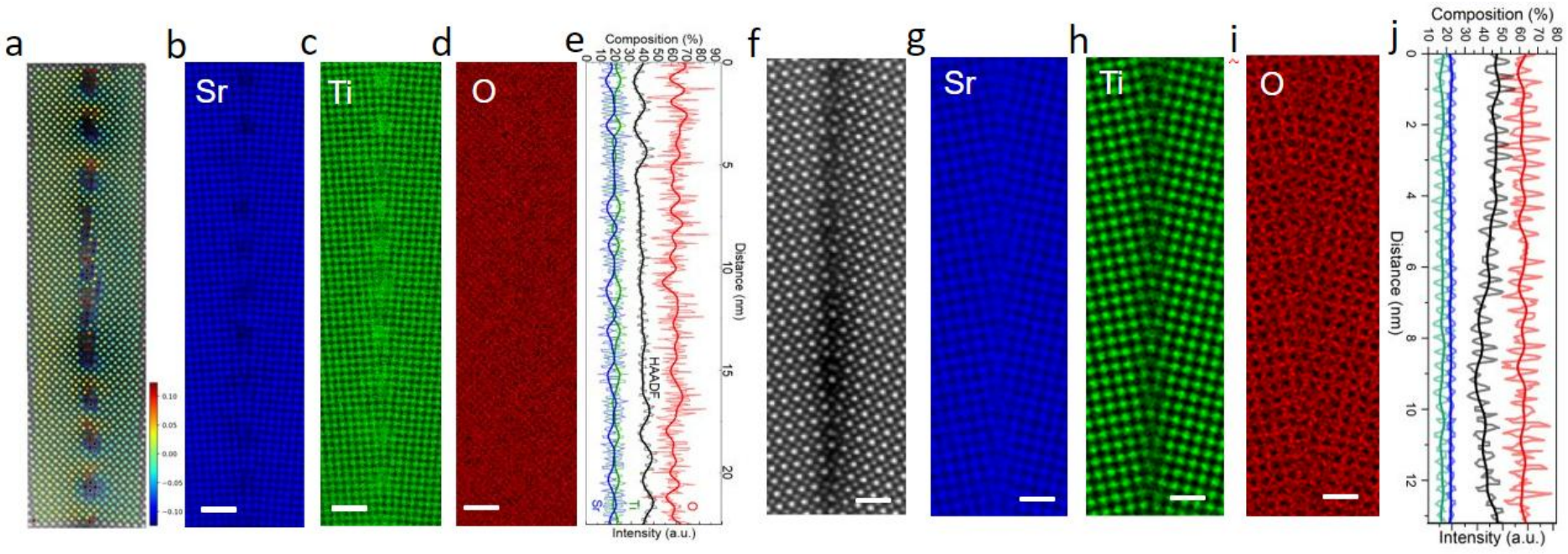

**Extended Data Fig. 4| EDS mapping.** (a) HAADF overlaid with strain mapping of 10° GB. Sr (b) , Ti (c), and O (d) mapping of 10° GB. (e) Line profile of HAADF, Sr, Ti, and O of 10° GB. (f) HAADF of 22°GB. Sr (g), Ti (h), and O (i) mapping of 22° GB. (j) Line profile of HAADF, Sr, Ti, and O of 22° GB.

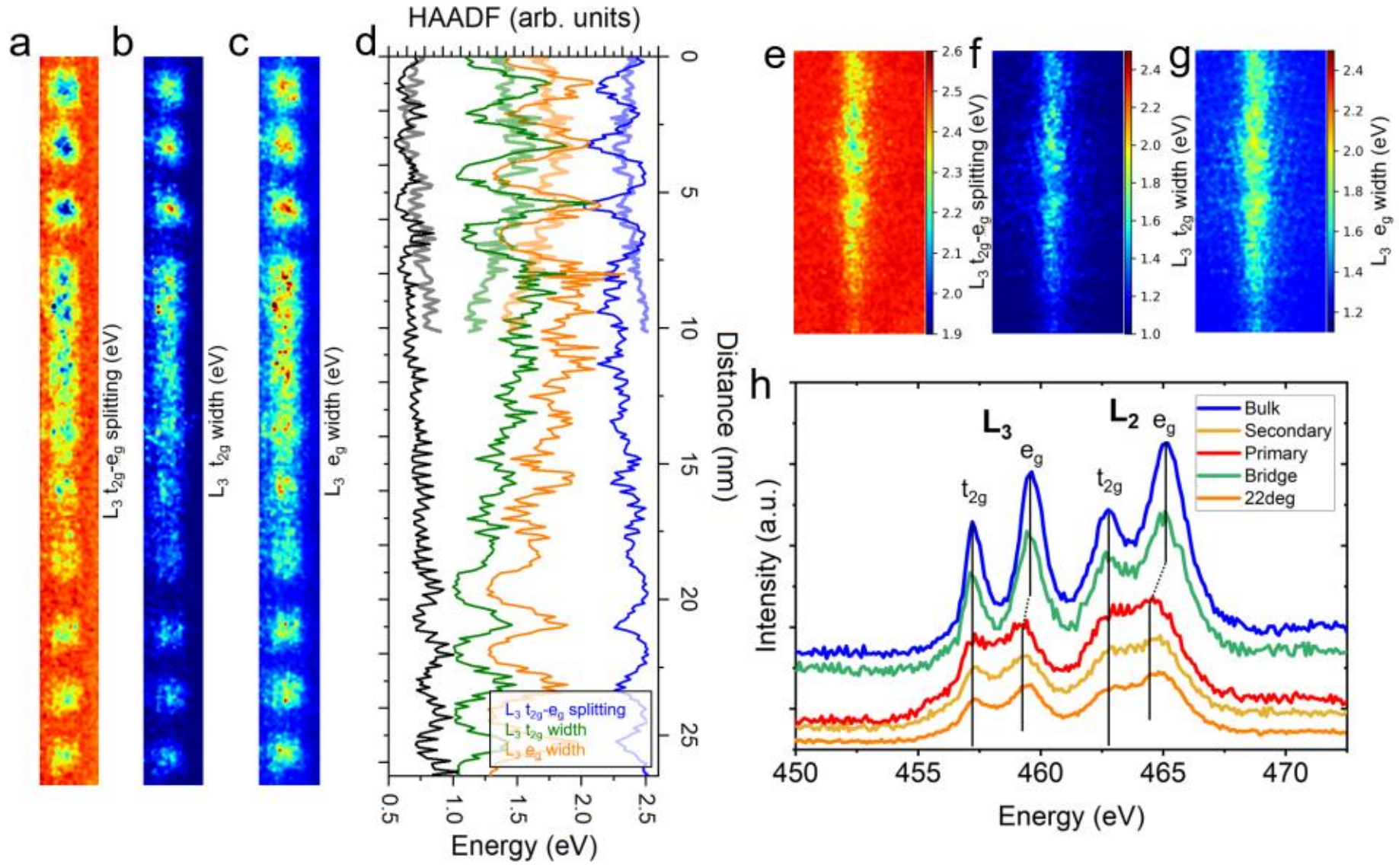

**Extended Data Fig. 5| Atomic strain, elemental map, and core loss investigation of bonding information.** (a) HAADF image with atoms colored by strain normal to GB plane. (b) Sr EDS mapping of 10° GB. (c) Ti mapping of 10° GB. (d) O mapping of 10° GB. Scale bar in **b-d** is 2nm. (e) Line profiles of Sr, Ti, O and HAADF. For clearer demonstration of trend FFT low-pass filters are overlaid in darker colors. (f) HAADF of 22° GB. (g) Sr EDS mapping of 22° GB. (h) Ti EDS mapping of 22° GB. (i) O EDS mapping of 22° GB. (j) Line profiles of Sr, Ti, O and HAADF of 22° GB. (k) $L_3$ $t_{2g}$-eg splitting of 10° GB. (l) $L_3$ $e_g$ width of 10° GB. (m) $L_3$-$t_{2g}$ width of 10° GB. (n) $L_3$ $t_{2g}$-eg splitting of 22° GB. (o) $L_3$ $e_g$ width of 22° GB. (p) $L_3$-$t_{2g}$ width of 22° GB. (q) Representative spectra of titanium L edge and O-K edge at bulk, tilt-induced defect core, and twist-induced core representative spectra of bulk, tilt and twist induced cores at the 10° and tilt induced defect cores at the 22° GB.

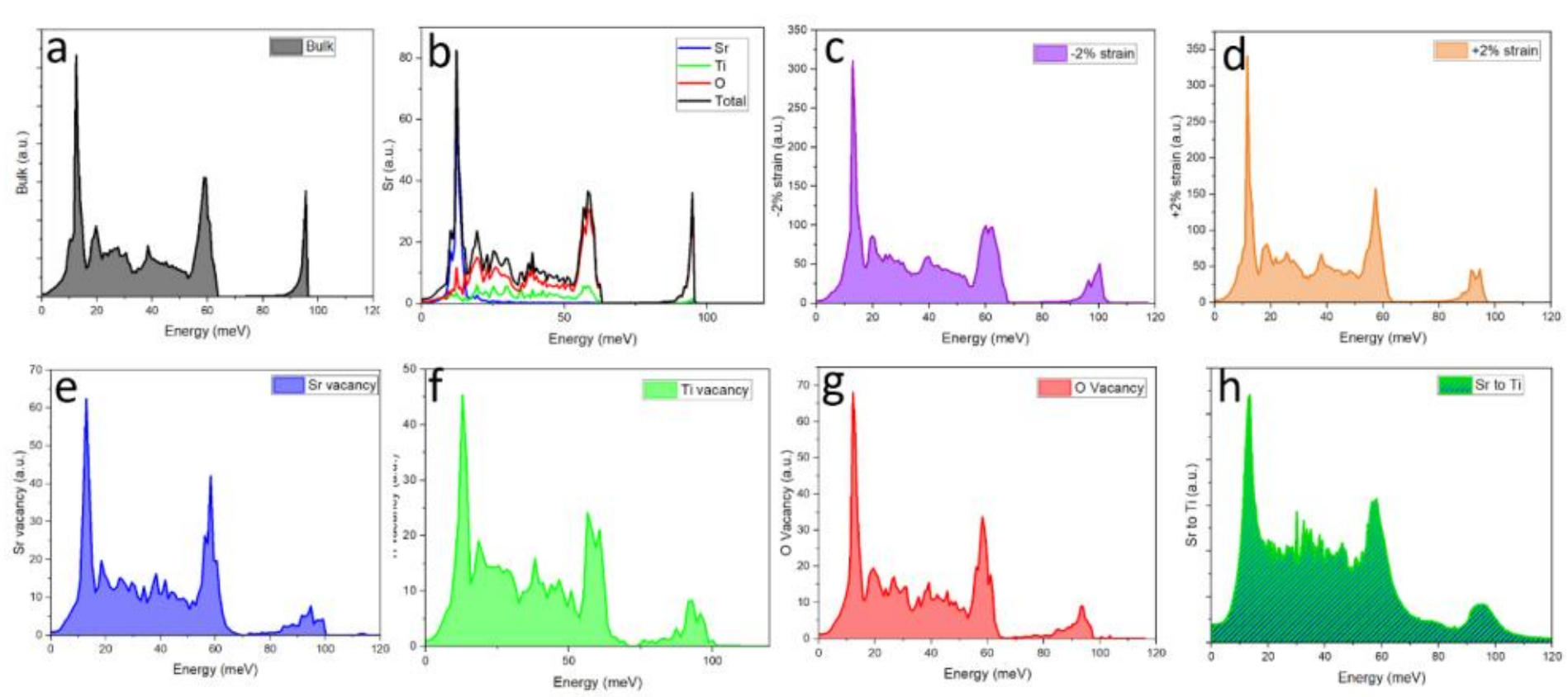

**Extended Data Fig. 6| DFT calculations.** (a) Unconvolved total PDOS of bulk STO. (b) Bulk PDOS with elemental contribution displayed. (c) Bulk PDOS with 2 percent compressive strain. (d) Bulk PDOS with 2 percent tensile strain. (e) PDOS with Sr vacancy. (f) PDOS with Ti vacancy. (g) PDOS with O vacancy. (h) PDOS with one Sr atom substituted by Ti.

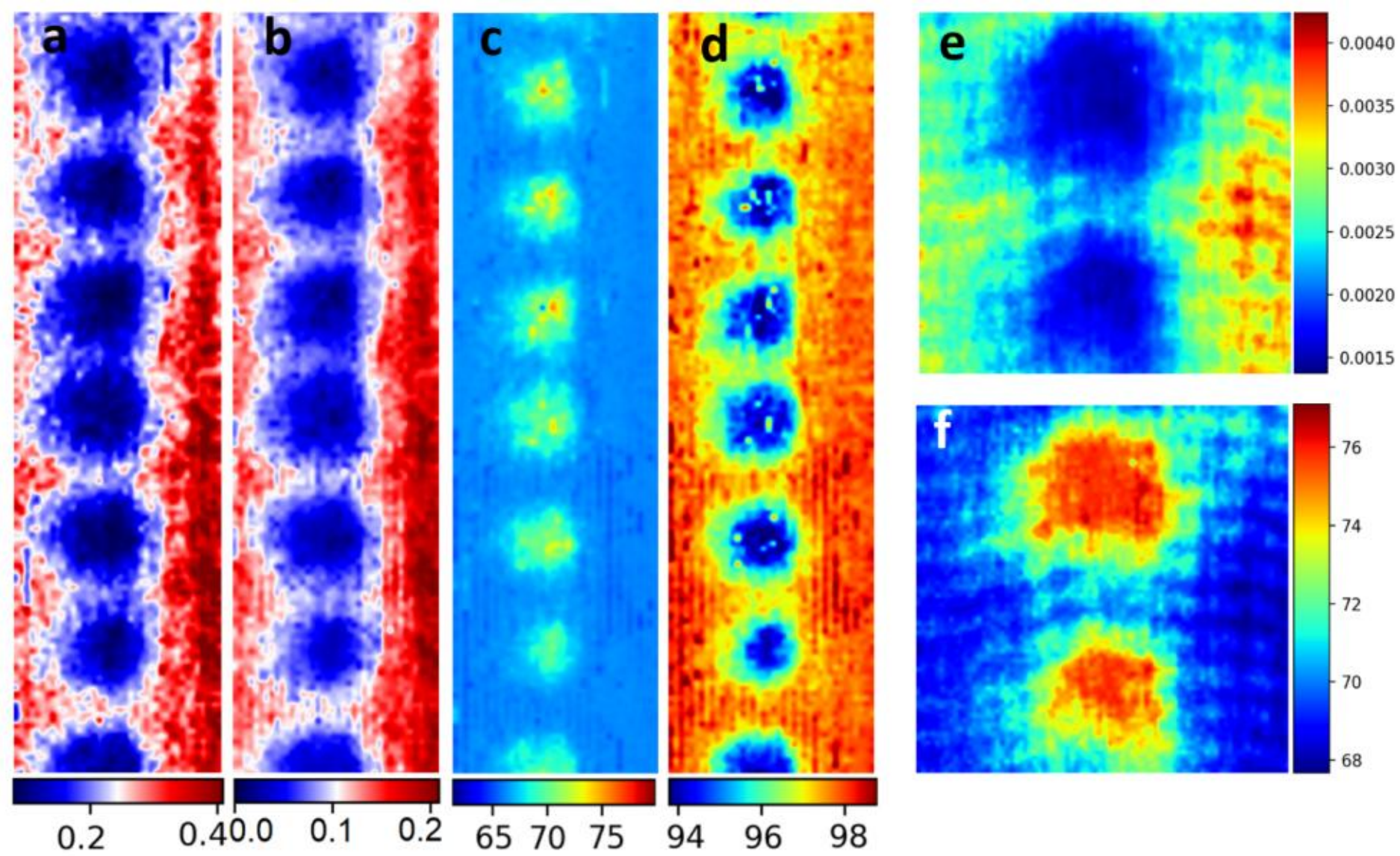

**Extended Data Fig. 7|Vibrational EELS mapping of tilt-induced cores of the 10° GB.** (a) LO2/TO3 area. (b) LO3 area. (c) LO2/TO3 position. (d) LO3 position. (e) High-magnification small step size LO2/TO3 height mapping of tilt-induced cores. (f) LO2/TO3 peak position mapping.

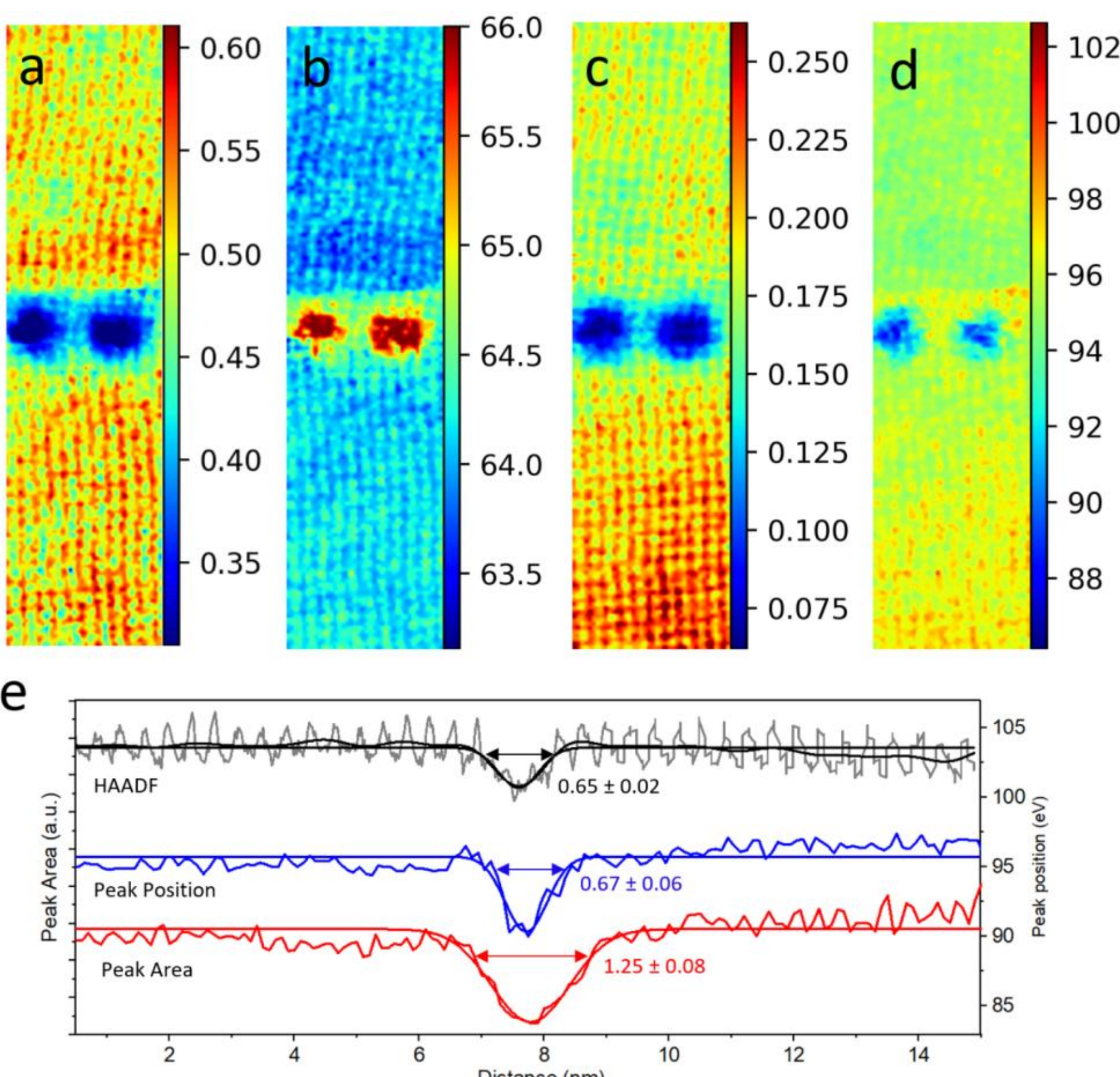

**Extended Data Fig. 8| Degree of signal localization visualization of the 10° GB.** (a) 60 meV peak area of two tilt-induced cores with mapping extended far into the bulk. (b) 50 meV peak position. (c) 100 meV peak area. (d) 100 meV peak position. (e) Line profiles of HAADF, 100 meV peak area, and 100 meV peak position. The degree of delocalization of HAADF and peak position is similar, but the peak area delocalization is visibly greater.

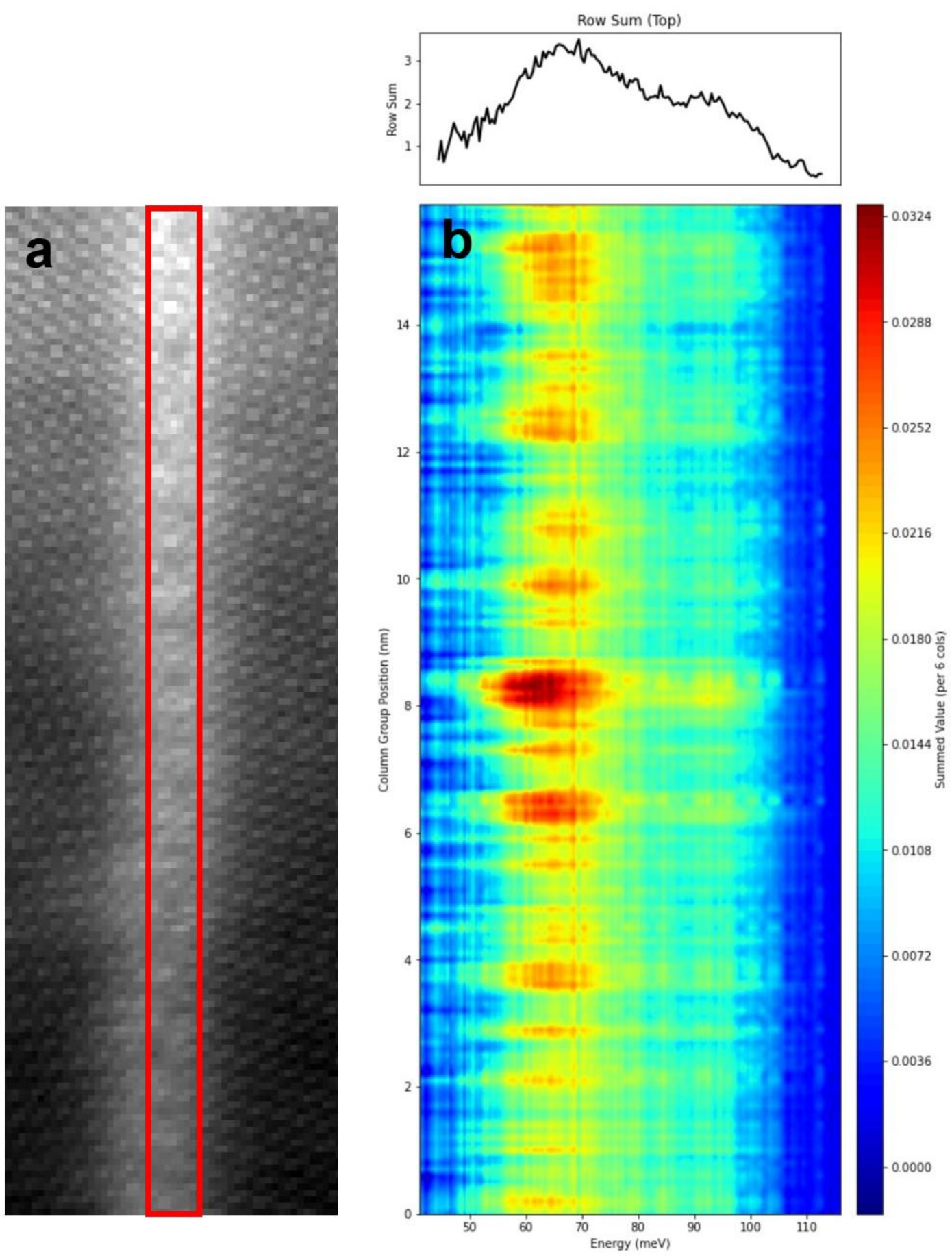

**Extended Data Fig. 9| Spectral contour of the 22° GB.** (a) 22 degree GB HAADF with GB line profile location indicated by red box. (b) Phonon spectral contour plot of the 22-degree GB.

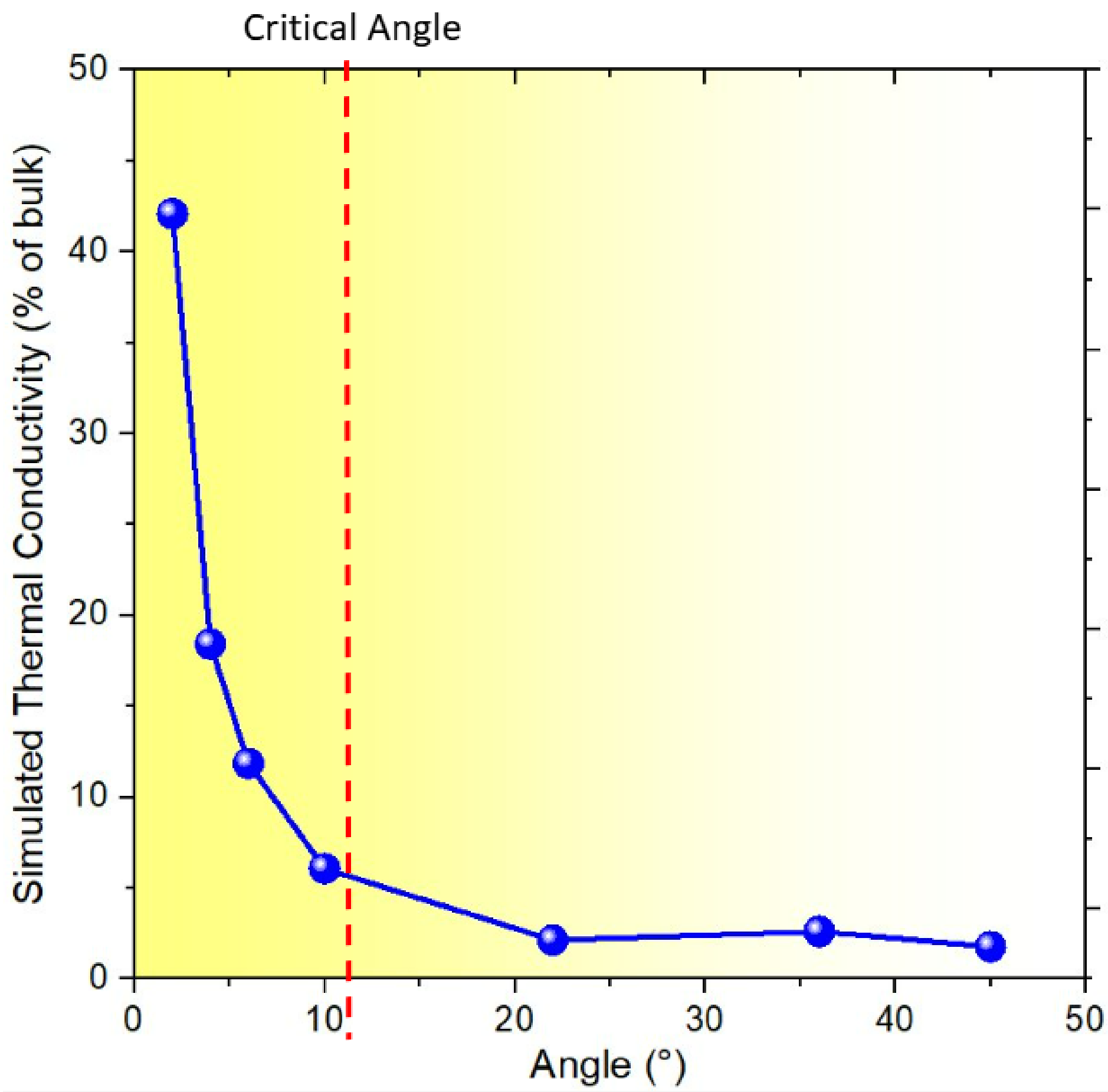

**Extended Data Fig. 10| simulated thermal conductivity vs tilt angle of the GBs.**

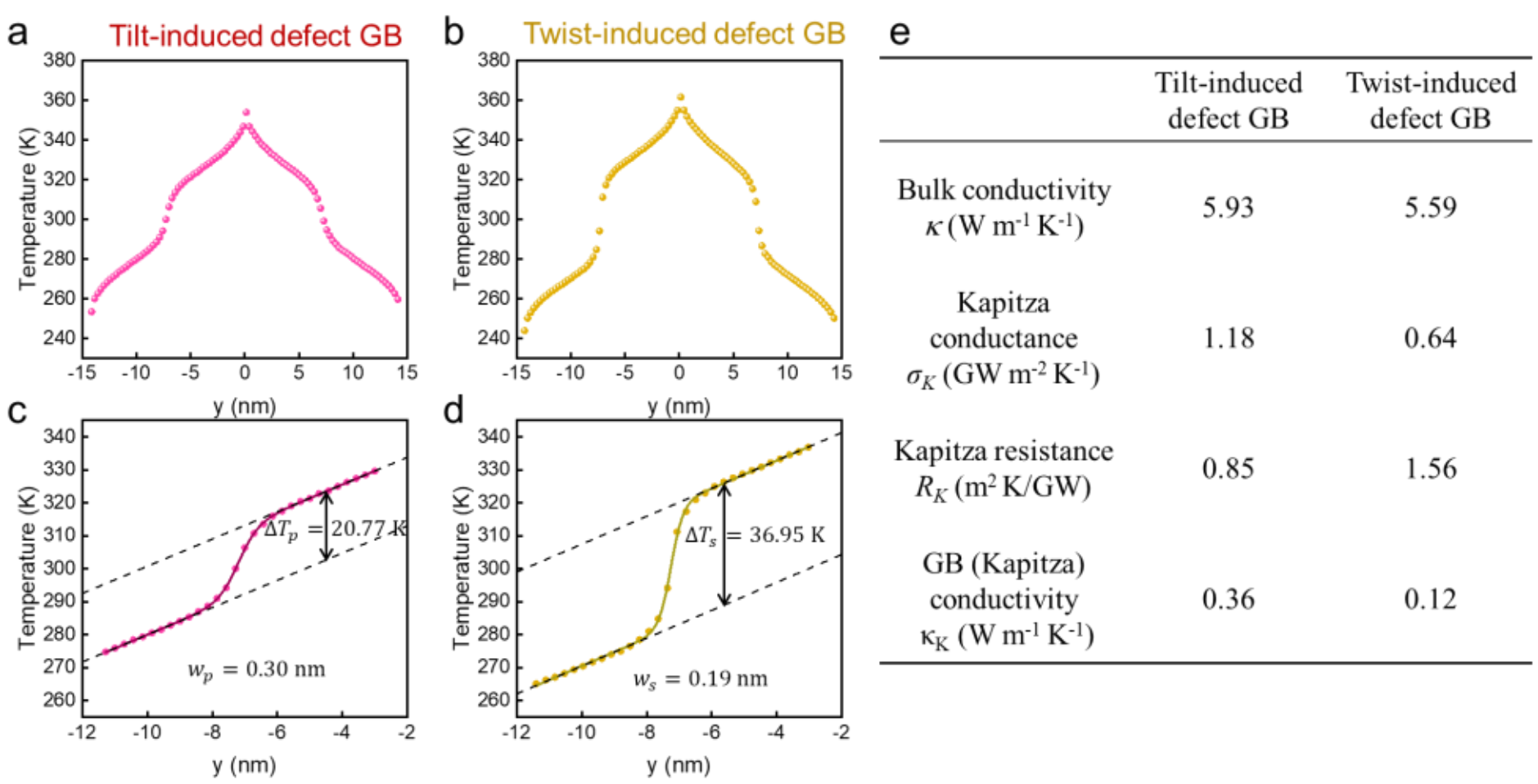

| | Tilt-induced defect GB | Twist-induced defect GB |
|---|---|---|
| Bulk conductivity $\kappa$ (W m$^{-1}$ K$^{-1}$) | 5.93 | 5.59 |
| Kapitza conductance $\sigma_K$ (GW m$^{-2}$ K$^{-1}$) | 1.18 | 0.64 |
| Kapitza resistance $R_K$ (m$^2$ K/GW) | 0.85 | 1.56 |
| GB (Kapitza) conductivity $\kappa_K$ (W m$^{-1}$ K$^{-1}$) | 0.36 | 0.12 |

**Extended Data Fig. 10| Thermal conductivity MD simulation of tilt and twist induced defect cores at the 10° GB.** (a)Temperature gradient of tilt-induced structure. (b)Temperature gradient of twist-induced structure. (c)Temperature change across the width of tilt-induced defect GB. (d)Temperature change across the width of twist-induced defect GB. (e)Table of simulated thermal conductivity of bulk, tilt-induced, and twist-induced defect GB.